\documentclass[journal]{IEEEtran}

\usepackage[pdftex]{graphicx}
\usepackage{epstopdf}
\ifCLASSINFOpdf
\else
\fi

\usepackage{xcolor}
\usepackage{ifthen}
\usepackage[normalem]{ulem}

\newcommand{\RISCV}{\mbox{RISC-V}}

\definecolor{darkgreen}{RGB}{0,128,0}

\definecolor{newtextcolor}{rgb}{0,0,0.5}

\definecolor{oldtextcolor}{rgb}{0.4,0.4,0.4}

\definecolor{revisedcolor}{RGB}{211,144,0} 
\newcounter{cntrevised}
\newcommand{\BEGINREVISED}{\refstepcounter{cntrevised}\color{revisedcolor}\raisebox{0.5ex}{\scriptsize\textbf{rev.\,\thecntrevised}}}
\newcommand{\ENDREVISED}{}

\usepackage{amssymb}
\usepackage{amsthm}
\usepackage{amsmath}

\theoremstyle{definition}
\newtheorem{definition}{Definition}

\usepackage{algorithm}
\usepackage[noend]{algpseudocode}

\makeatletter
\def\BState{\State\hskip-\ALG@thistlm}
\makeatother

\usepackage{hyperref}
\usepackage{tikz}

\newcommand\copyrighttext{\scriptsize %
  DOI: \href{https://doi.org/10.1109/TC.2022.3152666}{10.1109/TC.2022.3152666}. %
  \quad\textcopyright~2021 IEEE. Personal use of this material is permitted.
  Permission from IEEE must be obtained for all other uses, in any current or future
  media, including reprinting/republishing this material for advertising or promotional
  purposes, creating new collective works, for resale or redistribution to servers or
  lists, or reuse of any copyrighted component of this work in other works.
}

\newcommand\copyrightnotice{%
\begin{tikzpicture}[remember picture,overlay]
\node[anchor=south,yshift=7pt] at (current page.south) {\fbox{\parbox{\dimexpr\textwidth-\fboxsep-\fboxrule\relax}{\copyrighttext}}};
\end{tikzpicture}%
}

\begin{document}

%
%
\newcommand{\TitleFirst}{%
An Exhaustive Approach to Detecting 
}
%
\newcommand{\TitleSecond}{%
Transient Execution Side Channels
}
\newcommand{\TitleSird}{%
in RTL Designs of Processors
}
%
\title{%
\TitleFirst%
\\%
\TitleSecond%
\\%
\TitleSird%
}
\newcommand{\ExampleAuthor}{%
Mohammad~Rahmani~Fadiheh,
Alex~Wezel,
Johannes~M\"uller,
J\"org~Bormann,
Sayak Ray,
\\Jason M. Fung,
Subhasish~Mitra,~\IEEEmembership{Fellow,~IEEE}, 
Dominik~Stoffel, 
Wolfgang~Kunz,~\IEEEmembership{Fellow,~IEEE}
\thanks{ Mohammad Rahmani Fadiheh,
Alex Wezel,
Johannes M\"uller,
Dominik~Stoffel, and Wolfgang Kunz are with the Dept. of Electrical and Computer Engineering,
Technische Universit\"at Kaiserslautern, Germany.
}
\thanks{J\"org Bormann is with Siemens EDA, Germany.}
\thanks{Sayak Ray and Jason M. Fung are with Intel Product Assurance and Security, Intel Corporation, USA.}
\thanks{Subhasish Mitra is with the Dept. of Electrical and Computer Engineering and the Dept. of Computer Science,
Stanford University, USA.  }
\thanks{
The reported research was partly supported by DFG SPP Nano Security,
KU 1051/11-1, by BMBF ZuSE (Scale4Edge), 16ME0122K-16ME0124,
and by the Intel Corp. Side Channel Academic Program (SCAP)
}
}
%
\author{\ExampleAuthor}
%

%
%
%

%
%
%

%
%

\maketitle

\copyrightnotice

\begin{abstract}

  Hardware (HW) security issues have been emerging at an alarming rate
  in recent years. %
  Transient execution attacks, such as Spectre and Meltdown, in
  particular, pose a genuine threat to the security of modern
  computing systems. %
  Despite recent advances, understanding the intricate implications of
  microarchitectural design decisions on processor security remains a
  great challenge and has caused a number of update cycles in the
  past. %

  This papers addresses the need for a new approach to HW sign-off
  verification which guarantees the security of processors at the
  Register Transfer Level (RTL). %
  To this end, we introduce a formal definition of security with
  respect to transient execution attacks, formulated as a HW
  property. %
  We present a formal proof methodology based on \textit{Unique
    Program Execution Checking (UPEC)} which can be used to
  systematically detect all vulnerabilities to transient execution
  attacks in RTL designs. %
  UPEC does not exploit any a priori knowledge on known attacks and
  can therefore detect also vulnerabilities based on new, so far
  unknown, types of channels. %
  This is demonstrated by two new attack scenarios discovered in our
  experiments with UPEC. %

  UPEC scales to a wide range of HW designs, including in-order
  processors (RocketChip), pipelines with out-of-order writeback
  (Ariane), and processors with deep out-of-order speculative
  execution (BOOM). %
  To the best of our knowledge, UPEC is the first RTL verification
  technique that exhaustively covers transient execution side channels
  in processors of realistic complexity. %

\end{abstract}


\IEEEpeerreviewmaketitle

\section{Introduction}
\label{sec:introduction}

The discovery of Spectre~\cite{2018-KocherGenkin.etal} and
Meltdown~\cite{2018-LippSchwarz.etal} has led to a number of recent
product updates in industry, targeting both the hardware (HW) and the
software (SW) of computing systems. %
The variety of attacks discovered in the past years, from MDS attacks
(e.g., \cite{2019-CanellaGenkin.etal}) to Speculative
Interference~\cite{2020-BehniaSahu.etal}, with many of them attacking
a previously patched system (e.g., Fallout
attack~\cite{2019-CanellaGenkin.etal}), proves that transient
execution side channel (TES) attacks can pose a substantial security
challenge. %
TES attacks exploit side effects of transient instruction execution, a
phenomenon which is not visible in the sequential execution semantics
of the ISA. %
While transient instructions are not part of the correct flow of the
program, the processor still executes them but later discards their
result, e.g., due to a mispredicted branch. %

Software-level patches, such as in
~\cite{2020-OjogboThottethodi.etal}, are crucial for fixing legacy
systems, but they are not a panacea. %
Lack of concrete knowledge about the root cause of the problem can
lead to SW being patched based on the known attacks, potentially
rendering systems vulnerable to future attacks (%
see, e.g., the case of speculative store bypass
attack~\cite{2018-KirianskyWaldspurger}). %
Alternatively, SW may be fixed conservatively, preventing also future
attacks, however, at the cost of possibly prohibitive performance
overhead (up to~200\%~\cite{2020-BowenLupo}). %

  Compared to SW-only patches, enhancing security also at the HW level
  is the more favorable protection against TES attacks in the long
  run. %
  Besides improving the performance overhead~\cite{2019-YuYan.etal},
  microarchitectural countermeasures can %
  relieve the SW developer from having to consider HW details in
  security analysis. %

A variety of microarchitectural security patches have been proposed by
the research community. %
Although the achieved protection and the performance of these
techniques can be easily evaluated on abstract models (e.g., using the
gem5 simulator~\cite{2011-BinkertBeckmann.etal}), the security
\emph{guarantee} can, eventually, only be delivered at the Register
Transfer (RT)~level. %
As observed in the Speculative
Interference~\cite{2020-BehniaSahu.etal} and Spectre-STC attacks
(cf.~Sec.~\ref{sec:spectre-stc}), certain vulnerabilities are
introduced into the system due to specific Register Transfer Level
(RTL) details. %
In addition, the performance requirement for the microarchitectural
security patches leads to more and more complex design techniques
making the design prone to new vulnerabilities at the RTL. %

\subsection{Contributions}

We propose a formal security verification approach for HW, called
\emph{Unique Program Execution Checking (UPEC)}, which fully covers
the class of TES attacks. %
UPEC is applicable to RTL designs in a wide range of complexity, from
simple processors with in-order pipelines to deep out-of-order
speculative execution processors. %
Given a confidentiality requirement, UPEC either certifies the
security of the design w.r.t.\ this requirement, or it automatically
generates a counterexample pointing to possible vulnerabilities. %
The outline of the paper and its key contributions are as follows: %

(I) We provide new insights into TES attacks
(Sec.~\ref{sec:attacks}). %
Through experimental evidence, we show that TESs are possible not only
in high-end processors with advanced features, such as out-of-order
(OOO) and speculative execution. %
They can also exist in simple in-order processors and can be
introduced by low-level design decisions. %

(II) We propose formal definitions for microarchitectural
vulnerabilities causing confidentiality violation in HW through
explicit or implicit information flows
(Sec.~\ref{sec:upec-theoretical}). %
The definitions, unlike previous works, consider a concrete
microarchitecture on the RTL and do not restrict the leakage to known
side channels such as cache footprint. %
As a result, they also cover so far unknown TES
attacks. %

(III) This paper proposes UPEC as a structured and systematic formal
methodology for HW security verification targeting all TES attacks in
RTL processor designs (Secs.~\ref{sec:upec-bounded-model}
and~\ref{sec:upec-flow}). %
The proposed method is exhaustive and provides security guarantees
without requiring a radically different design methodology or expert
knowledge about different classes of microarchitectural attacks. %
UPEC therefore can be easily integrated into existing design flows and
can serve as a basis for HW security verification sign-off at the RT
level. %

(IV) The paper introduces the notion of \emph{microequivalence} which
is key in making UPEC scalable to processors with out-of-order and
speculative execution features (Sec.~\ref{sec:upec-for-ooo}). %

(V) The effectiveness of the UPEC approach is shown by verifying
various RTL designs with different levels of complexity
(Sec.~\ref{sec:experiments}). %
In our experiments, by merit of the exhaustiveness of our approach,
previously unknown vulnerabilities have been found in
RocketChip~\cite{2016-AsanovicAvizienis.etal},
Ariane~\cite{web-ariane} and BOOM~\cite{2017-CelioChiu.etal}. %
The reported runtime of UPEC confirms the scalability of the approach
to processors of realistic size. %

(VI) A sophisticated re-design of the BOOM processor is beyond the
scope of this paper. %
However, we developed a fully verified secure variant of the BOOM
processor based on conservative fixes, which is published as an open
source design on GitHub~\cite{web-upec-template}. %
Although the design does not deliver competitive performance, it
serves well as a demonstration that RTL sign-off verification
regarding TES attacks is doable for large processors. %
To the best of our knowledge, this is the first example of a deep OOO
processor with a well-defined security guarantee against transient
execution side channel attacks. %

\section{Related Work}
\label{sec:related-work}

Cheang et al.~\cite{2019-CheangRasmussen.etal} formally defined
\textit{Secure Speculation} as an observational
determinism~\cite{1995-Roscoe} property over four traces. %
A related approach was proposed by~\cite{2018-GuarnieriMorales.etal},
who formalized \textit{Speculative Non-Interference} as the
requirement for security against speculative execution attacks. %
These techniques can be effective in verifying software by extending
the ISA semantics to a speculative semantics. %
However, these and related works, such as
~\cite{2021-VassenaDisselkoen.etal, 2020-HeHu.etal}, do not take the
microarchitecture into account and are not meant to verify the
security of HW. %
Adopting SW security concepts to HW verification is not trivial, as
shown for the example of constant time execution
by~\cite{2019-GleissenthallKici.etal}. %

New HW/SW contracts have also been developed
in~\cite{2020-GuarnieriKoepf.etal} as a framework to reason more
precisely about what information HW leaks and what consequences this
has for SW security requirements. %
This provides a useful framework to reason about security at the SW
level, without leaving a gap due to certain microarchitectural
features. %
However, it has not yet been examined how the HW side of the
high-level contracts can be mapped to the RTL implementation by RTL
properties and whether these properties scale for state-of-the-art
commercial property checking. %

Cabodi et al.~\cite{2019-CabodiCamurati.etal} present one of the first
HW security verification approaches targeting speculative attacks. %
It extends over their previous work~\cite{2017-CabodiCamurati.etal}
and the work of~\cite{2014-SubramanyanArora}, pioneering the adoption
of \textit{taint analysis} in the HW domain. %
Hu et al.~\cite{2020-HuWu.etal} also proposed a model for precise
taint propagation and taint property verification. %
It is, however, in the nature of taint analysis that it requires
assumptions on the paths to be analyzed. %
As a result, such approaches can be effective in checking a design for
known variants of Spectre, but face their limitations with respect to
unknown variants. %
Furthermore, proving taint properties for paths of unknown temporal
length can be prohibitively complex, and restricting the proof to a
feasible temporal upper-bound limits the detection capability of the
method. %

The formal approaches known as
InSpectre~\cite{2020-GuancialeBalliu.etal},
CheckMate~\cite{2018b-TrippelLustig.etal} and the verification
techniques based on UCLID5~\cite{2018-SeshiaSubramanyan} evaluate the
security of HW designs on abstract models. %
  InSpectre~\cite{2020-GuancialeBalliu.etal} proposes an abstract
  execution semantics to formalize security against TES attacks. %
  This abstract formalization is different from the TES properties
  formulated in this paper, because it lacks the RTL semantics of the
  processor model. %
  Design verification based on high-level models %
  can provide important insights into the design. %
  However, it cannot provide the final sign-off verification which
  always uses cycle-accurate RTL as the point of reference. %

Also techniques based on HW fuzzing bear promise, such as
IntroSpectre~\cite{2021-GhaniyounBarber.etal} and
HyperFuzzing~\cite{2020-MuduliTakhar.etal}. %
Their non-exhaustive nature, however, does not allow them to deliver
formal guarantees when signing off a design. %

\section{New TES Attacks}
\label{sec:attacks}

\subsection{ORC Attack}
\label{sec:orc}

\newcommand{\ASSCODE}[1]{\texttt{\small #1}}
\newcommand{\REGX}[1]{\ASSCODE{x#1}}

For reasons of performance, many cache designs employ a pipelined
structure which allows the cache to receive new requests while still
processing previous ones. %
However, this can create a \emph{Read-After-Write (RAW) Hazard} in the
cache pipeline, if a load instruction tries to access an address for
which there is a pending write. %

A straightforward HW solution employs a \emph{hazard detection unit}
which checks for every read request whether or not there is a pending
write request to the same cache line. %
If so, all read requests are removed until the pending write has
completed. %
The processor pipeline is stalled, repeating to send read requests
until the cache interface accepts them. %

In the following, we show an example how such a cache structure can
create a security vulnerability allowing an attacker to open a TES. %
Let's assume we have a computing system with an in-order core
pipeline, a cache with write-back/write-allocate policy and the RAW
hazard resolution just described. %
In the system, some confidential data (\textit{secret data}) is stored
in a certain protected location (\textit{protected address}). %
We assume that the cache holds a valid copy of the secret data (from
an earlier execution of privileged code). %
We also make the simplifying assumption that each cache line holds a
single byte, and that a cache line is selected based on the lower 8
bits of the address of the cached location. %
Hence, in our example, there are a total of $2^{8}=256$ cache lines. %

\begin{figure}
  \begin{center}
  \ttfamily
  \scriptsize
  \begin{tabbing}
    XXXXX\=XXXXXXXXXXXXXXXXXXXXXXXXXX\=\kill
    \>  1:\' li x1, \#protected\_addr \>// x1 $\leftarrow$ \#protected\_addr \\
    \>  2:\' li x2, \#accessible\_addr \>// x2 $\leftarrow$ \#accessible\_addr \\
    \>  3:\' addi x2, x2, \#test\_value \>// x2 $\leftarrow$ x2 + \#test\_value\\
    \>  4:\' sw x3, 0(x2) \>// mem[x2+0] $\leftarrow$ x3 \\
    \>  5:\' lw x4, 0(x1) \>// x4 $\leftarrow$ mem[x1+0]\\
    \>  6:\' lw x5, 0(x4) \>// x5 $\leftarrow$ mem[x4+0]\\
  \end{tabbing}
  \caption{Example of an Orc attack (RISC-V code)}
  \label{fig:raw-attack}
  \vspace{-6ex} 
\end{center}
\end{figure}

The basic mechanism for the \emph{Orc attack} is the following. %
Every address in the computing system's address space is mapped to
some cache line. %
If we use the secret data as an address, then the secret data also
points to some cache line. %
The attacker program ``guesses'' which cache line the secret data
points to. %
It sets the conditions for a RAW hazard in the pipelined cache
interface by writing to the guessed cache line. %
If the guess was correct then the RAW hazard occurs, leading to a
slightly longer execution time of the instruction sequence than if the
guess was not correct. %
Instead of guessing, of course, the attacker program iteratively tries
all 256 possible cache locations until successful. %

Fig.~\ref{fig:raw-attack} shows the instruction sequence for one such
iteration. %
The sequence attempts an illegal memory access in instruction~\#5 by
trying to load the secret data from the protected address into
register~\REGX{4}. %
The processor correctly intercepts this attempt and raises an
exception. %
However, before control is actually transferred, instruction~\#6 has
already entered the pipeline and has initiated a cache transaction. %
The cache transaction has no effect on the architectural state of the
processor. %
But the execution time of the instruction sequence depends on the
state of the cache. %
When probing all values of \ASSCODE{\#test\_value}, the case will
occur where the read request affects the same cache line as the
pending write request of instruction~\#4, thus creating a RAW hazard and a stall in the
pipeline. %
It is this difference in timing that can be exploited as a side
channel. %

This new covert channel can be illustrated at the example of the
\RISCV{} Rocketchip~\cite{2016-AsanovicAvizienis.etal}. %
The original Rocketchip design is not vulnerable to the Orc attack. %
However, with only a slight modification (17 lines of code (LoC) in an
RTL design of $\sim$250,000~LoC) and without corrupting the
functionality, it is possible to insert the vulnerability. %
The modification actually optimizes the performance of the design by
bypassing a buffer in the cache such that an additional stall between
consecutive load instructions with data dependency is removed. %

The \emph{Orc attack} presented above is based on the interface between the core
(a simple in-order core in this case) and the cache. %
This provides the insight that the \textit{\underline{orc}hestration}
of component communication in an SoC, such as RAW hazard handling in
the core-to-cache interface, may also open or close covert
channels. %

The described vulnerability is a very subtle one, and unlike Meltdown
and Spectre, it is feasible in an in-order non-speculative pipeline. %

\subsection{Spectre-STC}
\label{sec:spectre-stc}

\newcommand{\SPECTREXYZ}{Spectre-STC}
\newcommand{\ATTACK}{\SPECTREXYZ}

\SPECTREXYZ{} is a new speculative transient execution attack which
demonstrates that \underline{s}ingle-\underline{t}hreaded processors
can be vulnerable to \underline{c}ontention-based Spectre attacks. %

In BOOM, the ALU, the multiplier and the division unit use the same
write port to access the register file. %
In case of contention, there is a fixed-priority arbitration, in which
the ALU has the highest and the division unit the lowest priority,
regardless of the program order. %
This means that a division instruction can be delayed by a younger
multiply or add instruction due to port contention, which is exploited
by \ATTACK{}. %

\begin{figure}
  \begin{center}
  \ttfamily
  \scriptsize
  \begin{tabbing}
    XXXXX\=XXXXXXXXXXXXXXXXXXXXXXXXX\=\kill
    \>  1:\' DIV x1,x2,x3 \>\textrm{// x1 $\leftarrow$ x2 $\div$ x3} \\
    \>  2:\' BR x1 $\ge$ 42, Line\_m \>\textrm{// if x1 $\ge$ 42 jump to line $m$} \\
    \>  3:\' LB x4, \textit{\textrm{addr\_of\_secret}} \>\textrm{// x4 $\leftarrow$ mem[\textit{addr\_of\_secret}]} \\
    \>  4:\' BR x4 $\neq$ \textit{\textrm{probe\_val}}, Line\_n \>\textrm{{// if x4 $\neq$ \textit{probe\_val} jump to line $n$}} \\
    \>  5:\' MUL \>\textrm{// a mult. instruction with arbitrary operands} \\
    \>  6:\' ADD \>\textrm{// an add. instruction  with arbitrary operands} \\
    \>    \' \ldots{} \>\textrm{// more such mult.'s and add.'s} \\
    \>  n:\' \ldots{} \>\textrm{// an arbitrary sequence of instructions\ldots} \\
    \>    \' \ldots{} \>\textrm{// \ldots other than \texttt{MUL} and \texttt{ADD}} \\
    \>  m:\' \ldots{} \>
    
  \end{tabbing}
	\caption{
		Pseudo-code of a \SPECTREXYZ{} gadget: %
    \textit{\textrm{addr\_of\_secret}} and
    \textit{\textrm{probe\_val}} are values 
		which can be controlled by attacker-provided inputs
	}

  \label{fig:attack-gadget}
  \vspace{-5ex}
\end{center}
\end{figure}

Fig.~\ref{fig:attack-gadget} shows a hypothetical gadget for
the \SPECTREXYZ{} attack. %
The basic idea is to start a division before the transient execution,
and then, in the transient time window, using a secret-dependent
branch, perform  multiplications and additions (and
create port contention) if the secret is equal to a probing value. %
The attacker can probe the secret (line 4) by measuring how long it takes to
execute the gadget code. %

This TES demonstrates that seemingly innocuous choices on RTL
implementation details, in this case on how to resolve the port
contention between two functional units, can introduce a vulnerability
to a Spectre-style attack. %

Similarly as \SPECTREXYZ{}, the research works on Speculative
Interference~\cite{2020-BehniaSahu.etal} and
SpectreRewind~\cite{2020-FustosBechtel.etal} discovered independently
that a secret-dependent branch instruction within the transient
sequence can be exploited as TES. %
In those attacks, the vulnerability is created by high-level
architectural features based on multi-cycle operations in
non-pipelined functional units. %
\SPECTREXYZ{}, discovered automatically by analyzing the RTL
description of BOOM, reveals that such vulnerabilities can have root
causes other than non-pipelined functional units. %

\section{Unique Program Execution Checking (UPEC)}
\label{sec:upec-theoretical}

In this section, we propose a mathematical definition for \emph{Unique
  Program Execution} and refine the formulation such that it
exclusively targets the class of TES attacks. %
Since UPEC shall be applicable to design implementations at the RTL,
all elements of the following definitions directly relate to
components of a CPU's microarchitecture. %

The contents of the program-visible registers and data memory
(including cache) are assumed to be partitioned into two sets: %
a set of public information, $X_p$ and a set of confidential
information, $X_c$. %
A program $P$ %
is the content of instruction memory together with a start address and
an end address. %
It receives $X_c$ and $X_p$ as its inputs. %
  In the microarchitectural implementation of our system, we consider
  a microarchitectural state~$S$, which is defined for all state bits
  other than those holding the inputs, $X_p$ and $X_c$, to the
  program. %
$S_0$~is the initial (starting) state for executing a program~$P$. %
We distinguish the following notions: %

\begin{definition}[Architectural Observation]
  \label{def:arch-observation}
  The \emph{architectural observation}~$O(P,X_p,X_c)$ of a program~$P$
  is the (time-abstract) sequence of valuations to the program-visible
  (architectural) registers, as they are produced by \emph{committing
    instructions} according to the ISA specification. %
  \hfill\qed%
\end{definition}

The architectural observation is independent of any hardware
implementation of the ISA and %
  does not relate to any microarchitectural features such as cache or
  pipelining. %
  It can be seen as the result of simulating the program by an ISA
  simulator. %
\begin{definition}[Microarchitectural Execution]
  \label{def:march-execution}
  The \emph{microarchitectural execution}~$\xi (S_0,P,X_p,X_c)$ of a
  program~$P$ is the (clock-cycle-accurate) sequence of valuations to
  the program-visible (architectural) registers, as they are produced
  during execution of~$P$ on a specific hardware
  implementation/microarchitecture. %
  \hfill\qed%
\end{definition}

The microarchitectural execution can be seen as the
clock-cycle-accurate result of simulating the program on the HW design
using an RTL simulator. %

\begin{definition}[Microarchitectural Observation] 
  \label{def:march-observation}
  The \emph{microarchitectural observation}~$\mu (S_0,P,X_p,X_c)$ of a
  program~$P$ is the (time-abstract) sequence of valuations to the
  program-visible (architectural) registers, as they are produced by
  \emph{committing instructions} on a specific hardware
  implementation/microarchitecture. %
  \hfill\qed%
\end{definition}

  In this definition, committing instructions are those instructions
  that finish their execution and leave the pipeline, i.e., they have
  written to the program-visible registers. %
The microarchitectural observation can be obtained from the
microarchitectural execution by keeping in the sequence all valuations
to the architectural registers at the time points they are written
(due to instruction commitment) and discarding all other
``intermediate'' valuations. %
For a functionally correct microarchitecture, for any tuple
$(P,X_p,X_c)$, the microarchitectural and architectural observation
are the same. %

Using these notions we can now formulate \emph{Unique Program
  Execution} which is the security requirement for an SoC to provide
the root of trust for confidentiality. %

\begin{definition}[Unique Program Execution]
  A program \emph{executes uniquely w.r.t.\ a set of confidential information~$X_c$} if and only
  if the sequence of valuations to the set of architectural state
  variables is independent of~$X_c$, in every clock cycle of program execution. %
  \hfill$\Box$ %
\end{definition}

We can use this notion to formulate the requirement for confidentiality as a trace property: %

\begin{align}
  \label{eq:upec-def} 
  \forall P, S_0, X_p, X_c, X_c': \nonumber \\ 
  O(P,X_p,X_c)&=O(P,X_p,X_c')  \nonumber \\
  \Rightarrow  \xi (S_0,P,X_p,X_c) &= \xi (S_0,P,X_p,X_c') 
\end{align}

The property is a ``non-interference property'' which considers an
arbitrary program running for different sets of secret data, $X_c$ and
$X_c'$. %
It states that if differences in the secret data do not change the
results of the program, as defined by the ISA level specification,
then its cycle-accurate execution for the program-visible registers of
the microarchitectural implementation must not depend on the secret
either. %

It is helpful to distinguish between confidentiality violations caused
by \emph{transient execution side channels (TESs)} and
\emph{functional leakages}. %
  The latter are not caused by timing side channels but by
  conventional design bugs violating the functional (ISA)
  specification of the system such that confidentiality is
  compromised. %
UPEC covers both types of confidentiality violations. %
A functional leakage is a counterexample not only to the property of
Eq.~\ref{eq:upec-def} but also to the following, weaker property: %
\begin{align}
  \label{eq:upec-functional-correctness} 
    \forall P, S_0, X_p, X_c, X_c': \nonumber \\ 
    O(P,X_p,X_c)&=O(P,X_p,X_c')  \nonumber \\
    \Rightarrow  \mu (S_0,P,X_p,X_c) &= \mu (S_0,P,X_p,X_c') 
\end{align}

This means, when targeting functional leakages, we do not consider the
detailed timing behavior of the HW implementation. %
We only need to check if the actual content of the program-visible
registers can be influenced by secret data. %
Functional leakages, in principle, can be identified by any functional
verification technique, such as conventional property checking or
functional simulation. %
Unfortunately, conventional verification techniques are not always
effective for this purpose. %
This motivates the work in~\cite{2021-MuellerFadiheh.etal} where the
potential of UPEC in detecting functional confidentiality violations
is examined further. %
This paper, however, is dedicated to TES-based confidentiality
violations. %

A \emph{TES attack} is a counterexample to the following
trace property. %
It results from Eq.~(\ref{eq:upec-def}) by excluding functional
leakages from consideration: %

\begin{align}
  \label{eq:transient-def}
  \forall P, S_0, X_p, X_c, X_c': \nonumber \\
  [\,\,O(P,X_p,X_c)&=O(P,X_p,X_c') \,\,\wedge \nonumber \\ 
  O(P,X_p,X_c) &= \mu (S_0,P,X_p,X_c) \,\,\wedge \nonumber \\ 
  O(P,X_p,X_c') &= \mu (S_0,P,X_p,X_c')\,\, ]  \nonumber \\ 
  \Rightarrow  \xi(S_0,P,X_p,X_c) &= \xi (S_0,P,X_p,X_c')
\end{align} 

It should be noted that this property verifies a concrete HW
considering any possible SW for both the attacker process and the
underlying operating system. %
  This is different from SW verification approaches
  (e.g.,~\cite{2019-CheangRasmussen.etal,2018-GuarnieriMorales.etal,2021-VassenaDisselkoen.etal,2020-HeHu.etal}) %
  which always consider a concrete kernel SW that must be verified to
  be secure against TES attacks. %

Consider a transient execution attack, given as a counterexample
$(S_0,P,X_p)$ to the trace property of Eq.~\ref{eq:transient-def},
executing %
  \emph{a concrete instruction sequence} $A$ %
  on the processor. %
  The attack is called a Spectre attack if there exists a sub-sequence
  of $A$ called $A_{victim}$ %
  such that $A_{victim}$ is executed under an execution context with %
  the proper privilege to access some confidential program
  input~$X_c$, without an exception being raised. %
  Otherwise, if no such $A_{victim}$ exists, the attack is called a
  Meltdown attack. %

  The counterexample to the above trace property is not necessarily
  the complete attack vector, but rather can be the critical part of
  the attack exfiltrating the secret and creating the covert
  channel. %
  For example, the part of the Spectre attack which poisons the branch
  prediction unit (by bad training) can be excluded from~$P$ and can
  be implicitly represented by~$S_0$. %
 
  Remember, $S_0$ defines the valuation to all state bits other than
  those holding $X_p$ and $X_c$. %
The use of the same~$S_0$ in both traces excludes classical timing
side channels from consideration, i.e., side channels which are not
based on transient executions. %
  Those rely on exploiting a specific victim SW, such as an encryption
  SW, whose footprint would need to be represented by a difference
  between initial states of the two traces. %

For the remainder of this paper, it is helpful to observe that any
valid counterexample to the property in Eq.~\ref{eq:transient-def}
fulfills the following condition: %
\begin{align}
\label{eq:micro-equivalence}
\mu(S_0,P,X_p,X_c) = \mu (S_0,P,X_p,X_c')
\end{align}

This condition is referred to as \textit{microequivalence} in the
following sections. %

\section{UPEC on a Bounded Model}
\label{sec:upec-bounded-model}

\newcommand{\PALERT}{\mbox{P-alert}}
\newcommand{\PALERTS}{\mbox{P-alerts}}
\newcommand{\LALERT}{\mbox{L-alert}}
\newcommand{\LALERTS}{\mbox{L-alerts}}
\newcommand{\CACHEMONITORVALIDIO}{\CODEVAR{cache\_monitor\_valid\_IO}}
\newcommand{\NOONGOINGPROTECTEDACCESS}{\CODEVAR{no\_ongoing\_protected\_access}}

\newcommand{\PUBLICMEM}[1]{\CODEVARSUB{public\_mem}{#1}} %

\newcommand{\CODEVAR}[1]{\mbox{\textrm{\textit{#1}}}}
\newcommand{\CODEVARSUB}[2]{\mbox{${\CODEVAR{#1}}_{#2}$}}
\newcommand{\SOC}[1]{\CODEVARSUB{SoC}{#1}}
\newcommand{\MEMORY}[1]{\CODEVARSUB{Memory}{#1}} %

\newcommand{\SECRETDATAPROTECTED}{\CODEVAR{secret\_data\_protected}}
\newcommand{\SECRETLOADSPECULATIVE}{\CODEVAR{secret\_load\_speculative}}
\newcommand{\SOCSTATE}[1]{\CODEVARSUB{soc\_state}{#1}}
\newcommand{\MICROSOCSTATE}[1]{\CODEVARSUB{micro\_soc\_state}{#1}}

Proving the property of Eq.~\ref{eq:upec-def} by generic unbounded
model checking is infeasible for SoCs of realistic size. %
Therefore, we develop a SAT-based approach which is tailored towards
the properties considered in UPEC. %
Our approach is based on ``\emph{any-state proofs}'' in a bounded
circuit model. %
It is related to a variant of Bounded Model Checking
(BMC)~\cite{1999-BiereCimatti.etal} called Interval Property Checking
(IPC)~\cite{2008-NguyenThalmaier.etal, 2014-UrdahlStoffel.etal} which
is applied to the UPEC problem in a similar way as
in~\cite{2018-FadihehUrdahl.etal} for functional processor
verification. %

The property of Eq.~\ref{eq:upec-def} relates two separate execution
traces for a given program and initial microarchitectural state %
to each other. %
Consequently, in our IPC-based approach, we use a computational model
that consists of two identical instances of the SoC under
verification, \SOC{1} and \SOC{2}, %
that each run an execution trace. %
In the terminology of SW verification, UPEC pursues an approach based
on \emph{2-safety hyperproperties}~\cite{2010-ClarksonSchneider}. %
A 2-safety formulation provides similar benefits for UPEC as for
information flow and taint property
verification~\cite{2017-CabodiCamurati.etal,2017-HuArdeshiricham.etal}. %
In HW equivalence checking the underlying self-compositional model is
widely referred to as ``miter''~\cite{1993a-Brand,
  2004-StoffelWedler.etal}. %
Owing to the resemblance of UPEC with HW equivalence checking we refer
to the UPEC computational model as ``miter for side channel
detection''. %

In the rest of the paper, without loss of generality, we assume that
$X_c$ resides in the data memory, unless otherwise noted. %
Verifying the SoC for confidentiality of information in the register
file is analogous. %

\subsection{UPEC Interval Property}
\label{sec:upec-interval-property}

\begin{figure}
  \centering
  \begin{minipage}{0.9\linewidth}
    \small
    \begin{tabbing}
      XXX\=XXXXXXXXXXX\=XXX\=XXX\=\kill
      \textcolor{blue}{assume:} \\
      \> \textcolor{blue}{at} $t$: \>$\MICROSOCSTATE{1} = \MICROSOCSTATE{2}$; \\
      \> \textcolor{blue}{at} $t$: \>$\PUBLICMEM{1} = \PUBLICMEM{2}$; \\
      \> \textcolor{blue}{at} $t$: \>\SECRETDATAPROTECTED{}(); \\
      \> \textcolor{blue}{during} $t$..$t+k$: \>\SECRETLOADSPECULATIVE{}(); \\
      \textcolor{blue}{prove:} \\
      \> \textcolor{blue}{at} $t+k$: \>$\SOCSTATE{1} = \SOCSTATE{2}$; \\
    \end{tabbing}
  \end{minipage}
  \caption{UPEC property formulated as interval property
  }
  \label{fig:ipc-property}
\end{figure}

\begin{figure}
  \centering
  \includegraphics[trim={96mm 27mm 75mm 25mm}, clip, width=\linewidth]{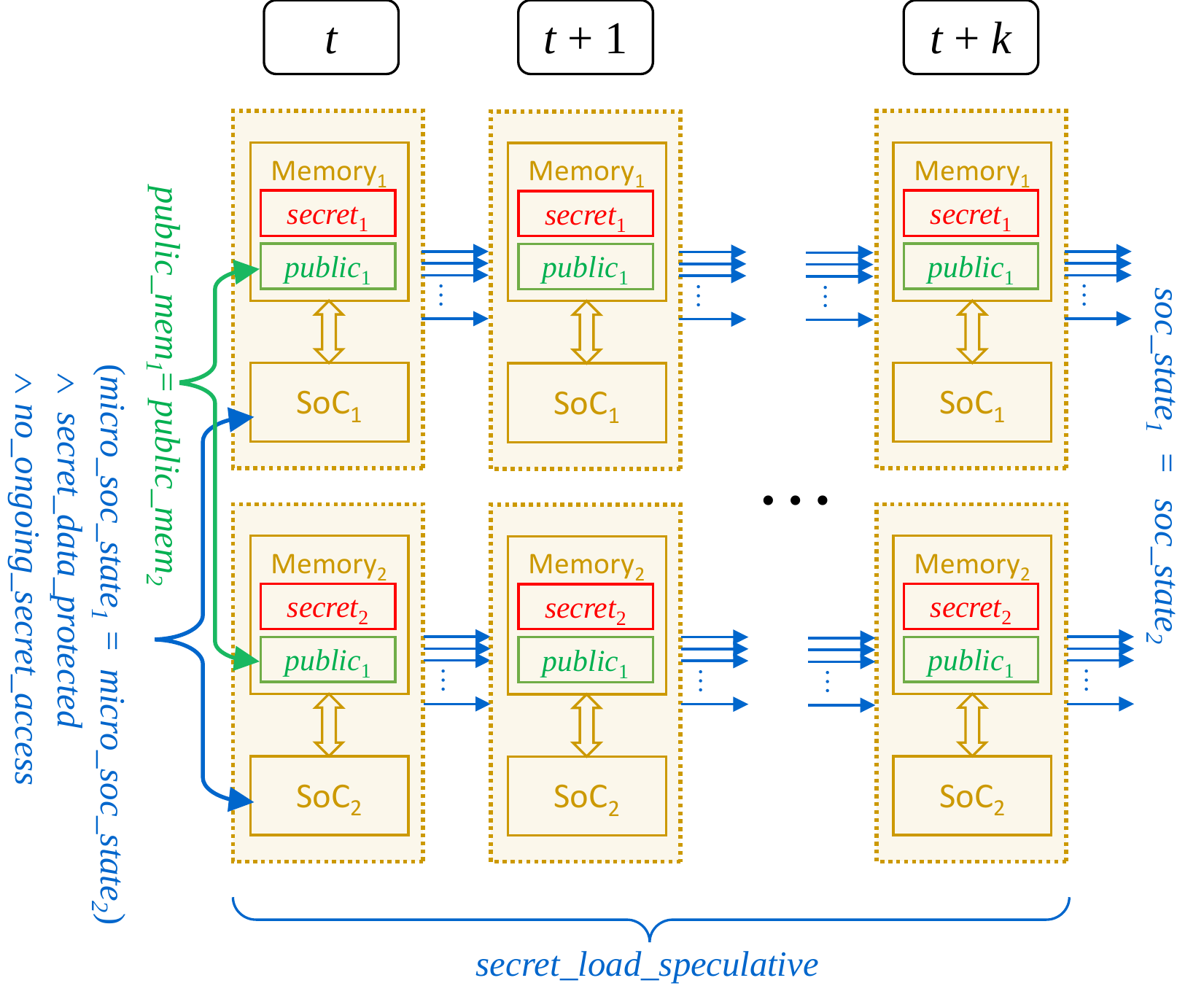} 
  \caption{Unrolled miter for side channel detection
  }
  \label{fig:computational-model-unrolled}
\end{figure}

Fig.~\ref{fig:ipc-property} shows the UPEC property formulated as an
interval property~\cite{2008-NguyenThalmaier.etal}. %
It is proven on the unrolled miter model depicted in
Fig.~\ref{fig:computational-model-unrolled}. %
In this property, \MICROSOCSTATE{} is the vector of all state
variables inside the SoC, excluding those parts of memory that hold
the public information~$X_P$ and the confidential information~$X_c$. %
The state variables of memory and cache holding $X_P$ are denoted with
\PUBLICMEM{} in the UPEC property. %

Similar to Eq.~\ref{eq:upec-def}, the first two assumptions of the
property guarantee that both SoC instances execute the same,
arbitrary, program and that the only difference between the two
execution traces at the start state is the set of secret data. %
   Since microarchitectural behavior at the RTL is
    deterministic, including that of cache eviction and branch
    prediction, %
  any discrepancy between the state of
  \SOC{1} and \SOC{2} that may occur in later clock cycles is due to
  the propagation of the secret. %

For a productive verification methodology, we need to avoid
formulating the property with reference to the architectural
observation of a program, since this would require formalizing the
ISA. %
Instead, we formulate two constraints \SECRETDATAPROTECTED{} and
\SECRETLOADSPECULATIVE{}. %
They are expressed in terms of the considered execution traces, and
serve the purpose of taking into account only traces with the same architectural
observation, without actually specifying the ISA. %

The macro \SECRETDATAPROTECTED{} states that a HW protection mechanism
is enabled to protect the memory locations holding~$X_c$. %
This means that, provided that the protection mechanism is implemented
correctly, any access to the secret within a user-level process is
blocked by an exception. %
This allows for considering Meltdown-style attacks (according to
Sec.~\ref{sec:upec-theoretical}). %
The macro \SECRETLOADSPECULATIVE{} assumes that any access to $X_c$
within an OS-level process happens transiently and is discarded before
it commits. %
This allows for considering Spectre-style attacks (according to
Sec.~\ref{sec:upec-theoretical}). %
  These macros can be easily refined for any given microarchitecture
  based on the memory protection logic and pipeline signals
  controlling the speculation level. %
  More detail on their implementation for the example of the BOOM
  processor can be found in~\cite{web-upec-template}. %
With the help of the macros \SECRETDATAPROTECTED{} and
\SECRETLOADSPECULATIVE{}, only execution traces are considered in
which any access to $X_c$ is either blocked by an exception or
discarded due to an earlier exception or misprediction. %
Since all other sources of instruction operands are the same between
the two instances, the two execution traces considered by the property
always yield the same architectural observation. %

Although the macros guarantee a secret-independent architectural
observation, it is important to understand that the described macros
do not over-constrain the interval property compared to the property
of Eq.~\ref{eq:upec-def}. %
In any execution trace that violates \SECRETDATAPROTECTED{} or
\SECRETLOADSPECULATIVE{}, the system executes a load instruction
targeting the secret that becomes visible as part of the architectural
observation, since the instruction is neither misspeculated nor
blocked by an exception. %
Consequently, the architectural observation is secret-dependent which
is a violation of the assumption in Eq.~\ref{eq:upec-def}. %
It should also be noted that \SECRETDATAPROTECTED{} and
\SECRETLOADSPECULATIVE{} macros do not provide a global constraint on
the system, but rather constrain program execution within the property
time window between $t$ and $t+k$. %
The system is still free to execute any program, even the ones
violating the macros, in the clock cycles before~$t$. %
This is taken into account by considering any microarchitectural state
as the starting state of the program execution for the trace property
of Eq.~\ref{eq:upec-def} and the interval property of
Fig.~\ref{fig:ipc-property}. %

As in any IPC proof, UPEC may return spurious counterexamples due to
considering unreachable starting states at the beginning of the
interval. %
This is addressed in the usual way by strengthening the property using
invariants. %
Since program execution is unique also in most unreachable states,
only little effort is caused in UPEC to create the needed invariants,
detailed in~\cite{2018-FadihehStoffel.etal}. %

\newcommand{\PROVE}{\textit{prove}}

  In the \PROVE{} part %
  of the property,
\SOCSTATE{} is a vector of state variables which includes all
architectural (program-visible) state variables, as required by
Eq.~\ref{eq:upec-def}, but additionally it can also include some or
all other microarchitectural state variables. %
Observing also non-architectural registers is a key concept in UPEC
which serves to obtaining an unbounded and scalable proof method. %
This will be further developed in the following subsections. %

\subsection{Counterexamples to the UPEC Property}
\label{sec:upec-counterexamples}

For a full proof of the UPEC property on the bounded computational
model of Fig.~\ref{fig:computational-model-unrolled}, in principle, we
need to consider a time window as large as the sequential depth,
$d_{soc}$, of the examined SoC. %
This is infeasible in most practical cases. %
Fortunately, employing a symbolic initial state allows the solver to
often capture hard-to-detect vulnerabilities within much smaller time
windows. %

Any information leakage starts with the propagation of confidential
information into some microarchitectural register. %
As a result, if we include every microarchitectural state variable in
\SOCSTATE{}, the any-state proof on our bounded model guarantees
detection of any violation of this UPEC property in only a single
clock cycle. %
This is due to the fact that the symbolic initial state of the proof %
includes, as a subset, all possible reachable states. %

However, unique program execution only requires that the secret data
must not be observable through the architectural state variables. %
Therefore, proving that the secret data cannot propagate to any state
variable in a superset of the architectural state variables represents
a sufficient but not a necessary condition for certifying
confidentiality. %

We therefore distinguish two kinds of ``alerts'' to the UPEC
property. %
In the following discussion, an alert~$a_k$ is a counterexample of
length~$k$ to the UPEC property where $\SOCSTATE{1} \neq\SOCSTATE{2}$
in the ending state, i.e., at time~$k$. %

\begin{definition}[\LALERT{}]
  \label{def:l-alert}
  \mbox{}\\
  An alert, $a_k$, is called a \emph{leakage alert
    (\LALERT{})} iff the differing state variables of \SOCSTATE{1} and
  \SOCSTATE{2} include \emph{architectural} state variables. %
  \hfill$\Box$%
\end{definition}

\LALERTS{} demonstrate how secret data can affect the sequence of
architectural states. %
Two cases can be distinguished. %
In the first case, secret data propagates directly into an
architectural register, i.e., the microarchitectural observation is
changed. %
If this happens the design contains a functional bug. %
In the second case, the problem is more subtle. %
The secret changes the timing and/or the values of the sequence
without violating the functional design correctness and without
leaking the secret directly. %
In other words, it changes the microarchitectural execution without
affecting the microarchitectural observation. %
UPEC detects the HW vulnerability in both cases. %

\begin{definition}[\PALERT{}]
  An alert, $a_k$, is called a \emph{propagation alert
    (\PALERT{})} iff the differing state variables of \SOCSTATE{1} and
  \SOCSTATE{2} are not \emph{architectural} state variables. %
  \hfill$\Box$ %
\end{definition}

\PALERT{}s show possible propagation paths of secret data from the
cache or memory to program-invisible, internal state variables of the
system, such as pipeline buffers. %
Hence, no secret data is \emph{leaked}. %
However, a \PALERT{} can be a precursor to an \LALERT{}, because the
secret often traverses internal, program-invisible buffers before it
is propagated to an architectural state variable like a register in
the register file. %
In fact, any \LALERT{} that is not (yet) visible within the time
window of the unrolled model has one or more shorter precursor
\PALERTS{} that are. %
Based on this observation, we realize that proving the absence of
\PALERTS{} and \LALERTS{} within one clock cycle in an any-state proof
is sufficient for verifying confidentiality. %
In the following section, we present the UPEC methodology for
verifying systems that do not fulfill this sufficient condition for
confidentiality. %

\section{UPEC Verification Methodology}
\label{sec:upec-flow}

\newcommand{\PPREDICATE}[1]{${\Lambda}_{#1}$}
\newcommand{\PVARS}[2]{\mbox{\text{\itshape Pvars\,(#1, #2)}}}
\newcommand{\PVAR}[1]{\PVARS{#1}{??}}
\newcommand{\PVARSPK}{\PVARS{$P$}{$k$}}
\newcommand{\PVARSNEW}[1]{\mbox{\text{\itshape Ploc\,(#1)}}}
\newcommand{\PVARSAK}{\PVARSNEW{$a_k$}}
\newcommand{\PNEXTVAR}[1]{$\mbox{Alert\_candidate}_{#1}$}
\newcommand{\PNEXTVARONE}[1]{${{Alert\_candidate}_{#1}}_{1}$}
\newcommand{\PNEXTVARTWO}[1]{${{Alert\_candidate}_{#1}}_{2}$}
\newcommand{\BC}[1]{\mbox{\textit{bc}$_{#1}$}}

\newcommand{\PLOC}{\mbox{P-location}}
\newcommand{\PLOCS}{\mbox{P-locations}}

\subsection{Modeling the Propagation of Secrets}
In the UPEC verification methodology, \PALERTS{} provide valuable
information on the ``locations'' to which secret information can
propagate. %

\begin{definition}[\PLOC{}]
  For a \PALERT{}~%
  $a_k$ %
  the \emph{\PLOC{}}, \PVARSAK, %
  is the set of all microarchitectural state variables of the SoC
  whose values in the ending state of $a_k$ (at time point $k$) depend
  on the secret, i.e., whose valuations differ between the SoC
  instances 1 and~2. %
  \qed
\end{definition}
In the following, we use \PLOCS{} to distinguish different
scenarios of secret information flows. %
In our proof method, the \PLOCS{} are represented by Boolean state
predicates. %

\begin{definition}[\PLOC{} Predicate]
  A \emph{\PLOC{} predicate}, \PPREDICATE{P}, for a set of
  microarchitectural state variables, $P$, is a state predicate over
  all microarchitectural variables of the miter, given as a
  conjunction of equalities and inequalities of corresponding variables
  in the SoC instances 1 and~2: %
  \begin{center}
    \PPREDICATE{P}$= \prod_i (v_{i,1} \neq v_{i,2}) \cdot \prod_j
    (w_{j,1} = w_{j,2})$
  \end{center}
  specifies
  \begin{itemize}
  \item inequality for every variable $v_i \in P$, and
  \item equality for every other variable $w_j \not\in P$.\qed%
  \end{itemize}
\end{definition}
A \PALERT{}~$a_k$ satisfies its \PLOC{} predicate~\PPREDICATE{P} with
$P = $~\PVARSAK{} at time point~$k$, i.e., in its ending state. %

Using these notions, we can collect and enumerate all possible
locations in the SoC to which secret information can propagate in~$k$
clock cycles. %
Similar to an all-SAT enumeration, after finding a P-alert~$a_k$, a
blocking clause is added to the property preventing the solver from
generating another counterexample fulfilling the same \PPREDICATE{P},
i.e., reaching the same \PLOC{}. %

In the following, we will use the complete set of \PLOCS{} to
decompose the UPEC proof problem into scalable sub-problems. %
Before this idea is elaborated, we introduce a cone-of-influence
reduction for UPEC, as an additional optimization for the proof
procedures to be presented. %

\subsection{Cone-of-Influence Reduction}
\label{sec:cone-of-influence-reduction}

Cone-of-influence reduction is a well-known technique for reducing the
complexity of proof tasks in formal verification by structurally
removing parts from the computational model that are irrelevant for
the proof at hand. %
In the original UPEC interval property of Fig.~\ref{fig:ipc-property},
however, the %
  \PROVE{} %
part includes every state variable in the design; %
as a result, the unrolled model cannot be simplified by the solver
through cone-of-influence reduction. %

Fortunately, this changes when we decompose the problem, as presented
in the following subsection. %
In the decomposed UPEC proof on a bounded model we are interested in
finding \PALERTS{} and \LALERTS{} of a specific length $k$ while the
precursor \PALERTS{} and their \PLOCS{} %
\PVARSNEW{$a_{k-1}$} %
are known. %
Therefore, only the subset of next-state variables that lie in the
fanout of the given locations need to be considered as the possible
candidates for a new alert. %

\begin{definition}[Alert Candidate]
    \textbf{\PNEXTVAR{P}} %
  is the set of state variables of the SoC which
  are in the immediate fanout cone of $P = $~\PVARSNEW{$a_{k-1}$}. %
\end{definition}

We can use the set~\PNEXTVAR{P} to simplify the proof obligation for
the property of Fig.~\ref{fig:ipc-property}. %
Assuming $a_{k-1}$ is a \PALERT{} of length $k-1$, for the UPEC proof
of length $k$, we can replace \SOCSTATE{} in the %
  \PROVE{} part of the property %
with \PNEXTVAR{P}. %
This helps the solver in many cases to conduct an effective
cone-of-influence reduction. %

\subsection{Iterative UPEC Proof Procedure}
\label{sec:upec-iterative}

The any-state proof of UPEC guarantees to find all \PALERTS{} of a
given temporal length, if the model is unrolled for the corresponding
number of clock cycles. %
As already observed in Sec.~\ref{sec:upec-counterexamples}, any
\PALERT{}~$a_k$ with $k > 1$ has a pre-cursor \PALERT{}~$a'_{k-1}$
which explains how the secret propagates to SoC variables in the
immediate fanin variables of \PVARSNEW{$a_k$}. %
This information can be used to guide the solver towards the next
counterexamples and mitigate the computational complexity of the proof
by decomposing the UPEC proof into simpler steps. %

In our general procedure, we enumerate \PALERTS{} by starting with the
ones for $k=1$, and then incrementing $k$ until no new alerts are
found. %
We decompose each UPEC proof for a particular~$k$ along the different
precursors for \PALERTS{}, by enumerating all \PLOCS{} reached within
$k-1$ clock cycles. %
The proposed decomposition does not violate the completeness of the
proof, since the any-state proof on the $k-1$ clock cycle time window
guarantees finding all possible \PALERTS{} of that length and, hence,
every counterexample to the UPEC property is covered by a \PLOC{}
computed by the proof procedure. %

\begin{figure}
  \centering
  \begin{minipage}{0.9\linewidth}
    \small
    \begin{tabbing}
      XX\=X\=XXXXXXXXXX\=XXX\=XXX\=\kill
      \textcolor{blue}{UPEC\_Base ($P$, $k$, blocking\_clause):}\\
      \> \textcolor{blue}{assume:} \\
      \> \> \textcolor{blue}{at} $t$: \>$\MICROSOCSTATE{1} = \MICROSOCSTATE{2}$; \\
      \> \> \textcolor{blue}{at} $t$: \>$\PUBLICMEM{1} = \PUBLICMEM{2}$; \\
      \> \> \textcolor{blue}{at} $t$: \>\SECRETDATAPROTECTED{}(); \\
      \> \> \textcolor{blue}{during} $t$..$t+k$: \>\SECRETLOADSPECULATIVE{}(); \\
      \> \> \textcolor{blue}{at} $t+k-1$: \>%
      \PPREDICATE{P} $=$ \textsf{\itshape\footnotesize  true}; \\
      \> \> \textcolor{blue}{at} $t+k$: \> blocking\_clause $=$ \textsf{\itshape\footnotesize  true}; \\

      \> \textcolor{blue}{prove:} \\
      \> \> \textcolor{blue}{at} $t+k$: \>%
      \PNEXTVARONE{P} = \PNEXTVARTWO{P}; \\
    \end{tabbing}
  \end{minipage}
  \caption{UPEC property for induction base procedure %
  }
  \label{fig:upec-property-parametric}
\end{figure}

Fig.~\ref{fig:upec-property-parametric} shows the UPEC property used
in our iterative procedure. %
The property receives a \PLOC{}~$P$, as reached by a
precursor \PALERT{}, and verifies whether the secret can propagate
further and produce a new alert. %
The property also receives a blocking clause which prevents the solver
from generating the same counterexamples again when calling the proof
procedure repeatedly. %

\begin{algorithm}
  \caption{UPEC\_Induction\_Base}\label{alg:upec-base}
  \begin{algorithmic}[1]
    \Procedure{UPEC\_induction\_base}{}
    \State $k \gets 1$
    \State $A_k \gets \emptyset$
    \State \BC{} $ \gets \textit{true}$
    \State $a_k \gets \text{IPC\_Solver}(\text{UPEC\_Base}(\emptyset, 1, \BC{}))$
    \While {$a_k \neq \emptyset$}
      \If {$a_k$ is \LALERT{}}
        \Return $a_k$
      \EndIf
      \State $P = $~\PVARSNEW{$a_k$}
      \State $A_k \gets A_k \cup $ \{P\}
      \State \BC{} $ \gets $ \BC{} $ \wedge \neg$ \PPREDICATE{P}
      \State $a_k \gets \text{IPC\_Solver}(\text{UPEC\_Base}(\emptyset, 1, \BC{}))$
    \EndWhile

    \While {$A_k \neq \emptyset$}
      \State $k \gets k+1$
      \State $A_k \gets \emptyset$
      \State \BC{} $ \gets \textit{true}$
      \For{$P'$ in $A_{k-1}$}
        \State $a_k \gets \text{IPC\_Solver}(\text{UPEC\_Base}(P', k, \BC{})$
        \While {$a_k \neq \emptyset$}
          \If {$a_k$ is \LALERT{}}
          \Return $a_k$
          \EndIf
          \State $P = $~\PVARSNEW{$a_k$}
    \If {$P \notin \bigcup_{i=1}^kA_i $}
      \State $A_k \gets A_k \cup $ \{P\}
    \EndIf
    \State \BC{} $ \gets $ \BC{} $ \wedge \neg$ \PPREDICATE{P}
    \State $a_k \gets \text{IPC\_Solver}(\text{UPEC\_Base}(P', k, \BC{}))$
        \EndWhile
      \EndFor
    \EndWhile
    \State \Return $\bigcup_{i=1}^kA_i$
    \EndProcedure
  \end{algorithmic}
\end{algorithm}

The UPEC verification methodology is based on proof by induction,
which consists of an induction base and an induction step. %
Alg.~\ref{alg:upec-base} shows the base proof. %
It iteratively verifies the design against the property of
Fig.~\ref{fig:upec-property-parametric}, collecting all possible
\PLOCS{}, and searches for any \LALERT{}. %
The ``IPC\_Solver'' is an interval property checker which returns a
counterexample if the given property fails or, otherwise, returns
$\emptyset$. %

The first step (lines 2 to~11) is to compute the initial set of
\PLOCS{}, $A_1$, which can be reached in one clock cycle. %
  ``UPEC\_Base($\emptyset$, 1, \BC{})'' %
  is a special case of the
UPEC property where $k=1$ and there is no predecessor \PALERT{}. %
Therefore, \PNEXTVAR{} needs to be determined based on the initial
location of the secret (e.g., data memory/data cache). %
The design is repeatedly verified using the property. %
Every time, after finding a counterexample, the blocking clause is
updated to search for new \PALERTS{}. %
The loop is terminated as soon as the solver returns $\emptyset$,
which means there is no new \PLOC{} that is reachable within one clock
cycle. %

Once $A_1$ is computed, the procedure continues to find more alerts
with longer time windows, until it reaches a time window in which it
determines no new \PALERT{} or \LALERT{} (lines 12 to~24). %
The \textit{for} loop (line~16) iterates through all \PLOCS{}
determined in the previous iteration for $k-1$, and, for each one of
them, computes all successor alerts of length $k$. %
In line~21, for the newly found alert $a_k$, we check whether the same
\PLOC{} has been detected by a shorter \PALERT{}. %
This avoids generating the same \PLOC{} twice and makes sure that
circular propagation of the secret between a set of state variables
will not cause the algorithm to become stuck in an infinite loop. %

Every time ``IPC\_Solver'' returns a counterexample, it is also
checked if it is an \LALERT{}. %
If this happens, a true security violation has been detected. %
The \LALERT{} is returned immediately to the designer for debugging
and repair. %
If no \LALERT{} is detected, the algorithm returns all the \PLOCS{}
found in the iterations. %

Once the design successfully passes the UPEC induction base proof, it is
guaranteed that, starting from \emph{any possible initial state}, there is no
information leakage possible within $k$~clock cycles. %
  The any-state proof within \emph{UPEC\_Induction\_Base} implicitly
  considers any possible pipeline context and system state that is
  required for initiating secret propagation. %
  As a result, for any information leakage possible within a time
  window larger than $k$~clock cycles, \emph{UPEC\_Induction\_Base} is
  guaranteed to find at least one precursor \PALERT{}. %
  This ensures exhaustive coverage of yet undiscovered \PALERTS{} and
  \LALERTS{} by the following algorithm for the induction step. %

The goal of the induction step is to prove that, if the secret
propagates to some non-observable microarchitectural state variables,
it will not leak to architectural state variables at any time in the
future. %
In other words, we need to prove that a \PALERT{} cannot be extended
further to reach a new \PLOC{} or to an \LALERT{}. %
Therefore, in contrast to the induction base, the induction step
property does not assume equality of all microarchitectural state
variables at time $t$ but instead assumes that the secret has already
propagated to some of these variables, represented by a
\PLOC{}~$P'$. %

\begin{figure}
  \centering \small
    \begin{tabbing}
      XX\=X\=XXXXXXXXXX\=XXX\=XXX\=\kill
      \textcolor{blue}{UPEC\_Step ($P$, blocking\_clause):}\\
      \> \textcolor{blue}{assume:} \\
      \> \> \textcolor{blue}{at} $t$: \>$\PUBLICMEM{1} = \PUBLICMEM{2}$; \\
      \> \> \textcolor{blue}{at} $t$: \>\SECRETDATAPROTECTED{}(); \\
      \> \> \textcolor{blue}{during} $t$..$t+2$: \>\SECRETLOADSPECULATIVE{}(); \\
      \> \> \textcolor{blue}{at} $t$: \>\PPREDICATE{P} $=$ \textsf{\itshape\footnotesize  true}\\
      \> \> \textcolor{blue}{at} $t+1$: \>\PNEXTVARONE{P} $=$ \PNEXTVARTWO{P}; \\
      \> \> \textcolor{blue}{at} $t+2$: \>blocking\_clause $=$ \textsf{\itshape\footnotesize  true};
      \\
      \> \textcolor{blue}{prove:} \\
      \> \> \textcolor{blue}{at} $t+2$: \>\PNEXTVARONE{P} $=$ \PNEXTVARTWO{P}; \\
    \end{tabbing}
    \vspace{-4ex}
  \caption{UPEC property for induction step procedure}
  \label{fig:upec-inductive-step}
\end{figure}

  \begin{algorithm}
  \caption{UPEC\_Induction\_Step}\label{alg:upec-step}
  \begin{algorithmic}[1]
    \Procedure{UPEC\_induction\_step}{$A$}
  \For {$P'$ in $A$}
    \State \BC{}$ \gets \textit{true}$
    \State $a_2 \gets \text{IPC\_Solver}(\text{UPEC\_Step}(P', \BC{}))$
    \While {$a_2 \neq \emptyset$}
      \If {$a_2$ is \LALERT{}}
        \Return $a_2$
      \EndIf
      \State $P = $~\PVARSNEW{$a_2$}
      \If { $P \notin A $}
        \State $A \gets A \cup \{P\}$
      \EndIf
      \State \BC{} $ \gets $ \BC{} $ \wedge \neg$ \PPREDICATE{P}
      \State $a_2 \gets \text{IPC\_Solver}(\text{UPEC\_Step}(P', \BC{}))$
    \EndWhile
  \EndFor
      \State \Return $\emptyset$
    \EndProcedure
  \end{algorithmic}
\end{algorithm}
Fig.~\ref{fig:upec-inductive-step} shows the property which is the
heart of \emph{UPEC\_Induction\_Step} (Alg.~\ref{alg:upec-step}). %
In this property, the starting state (at time point $t$) is the ending
state of some \PALERT{} found within the base proof. %
The property proves that any \PALERT{} (of any length) that reaches
$P$ and that does not reach a new \PLOC{} in the subsequent clock
cycle, will also (inductively) not reach a new \PLOC{} thereafter. %
Note that a counterexample to the step property is always of
length~2. %

The procedure over-approximates the history prior to the ending state
of a considered \PALERT{} and, thus, includes all reachable behaviors
of the design, including possible \LALERTS{}. %
The over-approximation may lead to unreachable counterexamples. %
This is a standard problem and can be addressed by extending the proof
to a $k$-step induction~\cite{2000-SheeranSingh.etal}, or by
strengthening the initial state of the proof by invariants. %
In our experiments, only for one of the designs (RocketChip), due to
several uninitialized state variables, a few simple invariants had to
be added manually. %

Alg.~\ref{alg:upec-step} describes the procedure for the UPEC
induction step. %
It receives the set of \PLOCS{} determined by the induction base
computation of Alg.~\ref{alg:upec-base}. %
For each \PLOC{} in this set, it computes all subsequent \PALERTS{}
using the ``UPEC\_Step'' property. %
If the property holds, it is guaranteed that there is no subsequent
\PALERT{}. %
If the property fails, it is checked whether the counterexample is an
\LALERT{}. %
If it is not an \LALERT{} and also yields a new \PLOC{}, the new
\PLOC{} is added to the set%
   (lines 8 and~9) %
and used in another ``UPEC\_Step'' proof. %

  The latter situation reflects that the induction base computation of
  Alg.~\ref{alg:upec-base} terminates as soon as incrementing~$k$ does
  not yield any new \PLOC{}. %
  The bounded model of \emph{UPEC\_Induction\_Base} misses a \PLOC{} in case it is
    reached only at a later time point, after having been propagated through already known \PLOCS{}. %
  However, such a \PLOC{} is detected afterwards, in
    \emph{UPEC\_Induction\_Step}, by iteratively computing all
    successor \PLOCS{} of already known ones. %
    Alg.~\ref{alg:upec-step}, therefore, always completes the set of
    \PLOCS{}. %
  If Alg.~\ref{alg:upec-step} terminates without an
  \LALERT{}, it means there cannot be any information leakage and the
  procedure returns~$\emptyset$, certifying the security of the
  design. %

In most practical cases, already Alg.~\ref{alg:upec-base} determines
all \PLOCS{} that exist. %
This implies that \emph{UPEC\_Induction\_Step} mostly serves as a check for
proof completeness and usually does not incur additional debugging
effort. %
  The possible \PLOCS{} typically comprise only a small fraction of
  the microarchitectural state variables, entailing only a small
  number of iterations in Alg.~\ref{alg:upec-base} and Alg.~\ref{alg:upec-step}. %
Although the above procedures are fully automated, there can be a
benefit in manually inspecting the counterexamples for every generated
\PALERT{} that points to a new \PLOC{}. %
This helps the user to understand the security implications of
microarchitectural optimizations, detect undesired information flows
and, even more importantly, capture security violations early if the
security compromise is already obvious from the given \PALERT{}. %

\subsection{Blackboxing in the UPEC Flow}

The computational model of UPEC provides a good opportunity for
automatic blackboxing, without the risks associated with under- or
over-approximating the design behavior. %
In UPEC a component can be blackboxed soundly using only a simple
constraint. %
It states that the component outputs are equal between SoC instance 1
and~2 in every clock cycle of the proof as long as the component's
inputs are equal between SoC instance 1 and~2 in every clock cycle of
the proof. %
If a \PALERT{} occurs during the verification procedure of
Sec~\ref{sec:upec-iterative} which violates this condition, i.e., if
the secret propagates to the inputs of the blackboxed component, then
this component must be unblackboxed. %
Otherwise, it is proven that the blackbox is sound and no leakage can
be missed. %

This blackboxing technique can, for example, be applied effectively to
system components containing large memory arrays, like the tag array
in a cache. %
It should be noted that in the design examples we have considered so
far, such components have a state-dependent timing behavior. %
Consequently, secret propagation to their internal state can alter the
timing of instruction execution. %
Therefore, a \PALERT{} which shows a propagation of the secret to the
inputs of such a component can be seen as an early security alarm. %

\section{UPEC for Advanced Processors}
\label{sec:upec-for-ooo}

\subsection{UPEC for OOO processors}
\label{sec:upec-for-ooo-proc}

For processors of medium complexity with \textit{in-order} pipelining,
the miter construction of Fig.~\ref{fig:computational-model-unrolled},
as discussed in Sec.~\ref{sec:upec-interval-property}, entails that
only little effort is required for manually creating invariants to
eliminate unreachable
counterexamples~\cite{2019-FadihehStoffel.etal}. %
However, this changes when out-of-order processors are considered. %
The symbolic initial state can then include starting states, for
example, with inconsistent instruction tags or IDs in different
bookkeeping structures, so that invalid execution orders are
considered. %

The key idea to tackle this problem is to constrain the UPEC proof by
\textit{microequivalence} (Eq.~\ref{eq:micro-equivalence}). %
This means that, instead of verifying the trace property of
Eq.~\ref{eq:upec-def}, we propose to target Eq.~\ref{eq:transient-def}
when dealing with OOO-processors. %
This simplifies the proof procedure significantly. %
Note that excluding functional correctness violations from
consideration is legitimate since functional verification%
  , i.e., checking compliance of the microarchitectural implementation
  with the functional ISA specification, %
  is a separate and %
  mandatory step in standard design flows, regardless of the security
  requirements. %

In the following, we provide an exemplary description of how to
specify microequivalence. %
A general template for microequivalence and a concrete example of its
refinement for the BOOM processor is made available
at~\cite{web-upec-template}. %

The symbolic initial state models the reservation stations of the
pipeline with arbitrary instructions having arbitrary operands, IDs
and other bookkeeping information. %
The information about how this is linked by the renaming mechanism to
the original instructions, as they were processed by the previous
decode stage, is lost. %
For example, in our computational model, an older, non-speculative ADD
instruction may receive an operand from the destination register of a
younger speculative LOAD instruction. %
Obviously, this can lead to false alarms. %

Invariants for microequivalence can be created that remove such
spurious behavior by relating the entries of the reservation stations
to the relevant operand and bookkeeping information of the reorder
buffer (ROB) such that only feasible orders and operand dependencies
are taken into consideration. %
These invariants must be formulated, however, for each pair of pending
instructions in the pipeline. %
This, unfortunately, can be quite laborious. %
 
\begin{figure}
  \centering
  \includegraphics[trim={4mm 102mm 184mm 16mm}, clip, width=0.6\linewidth]{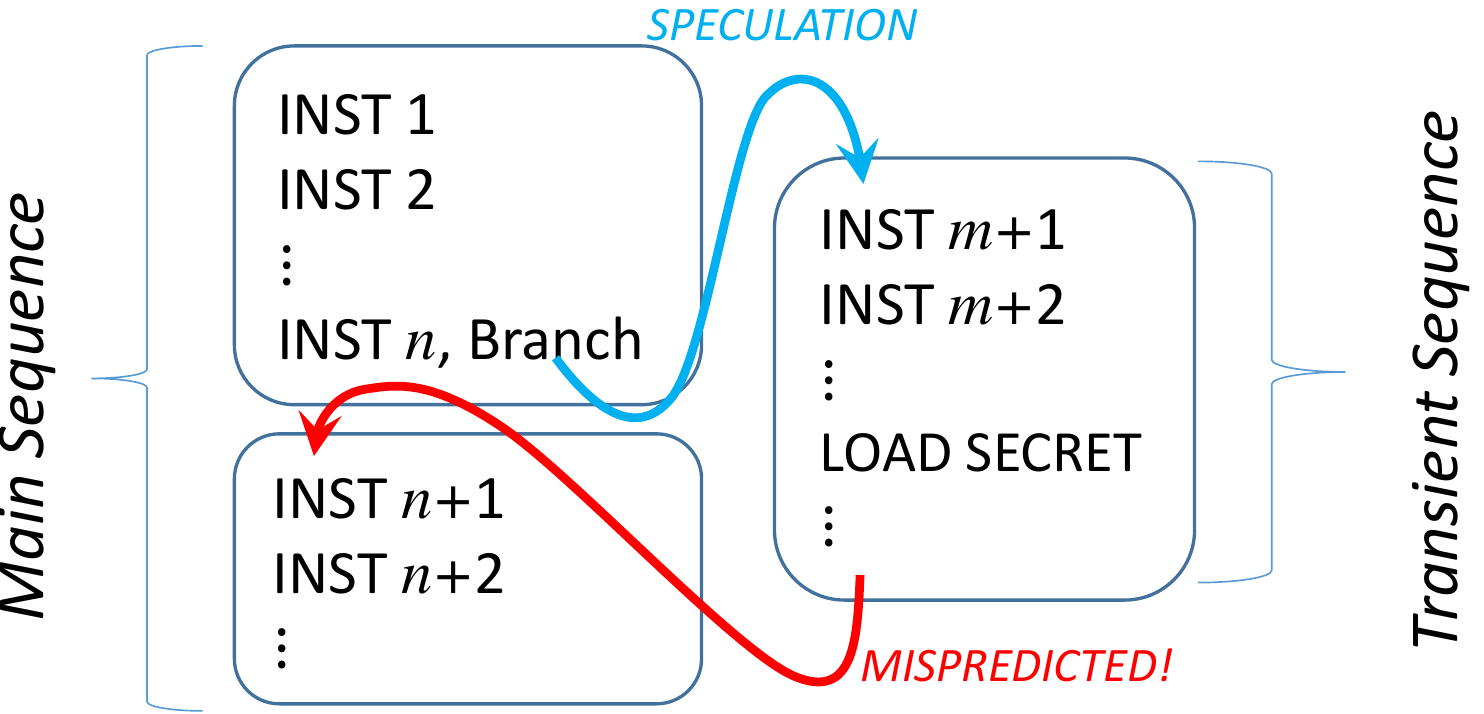} 
  \caption{General model for transient execution attack: %
    microarchitectural flow}
  \label{fig:transient-inst-seq}
  \vspace{-3ex}
\end{figure}

A better understanding of transient instruction execution in OOO
processors helps us to to reduce this manual effort drastically. %
Fig.~\ref{fig:transient-inst-seq} shows a general model of how
instructions execute in a transient execution attack. %
The \emph{main sequence} of instructions reaches a branch instruction,
initiating the \emph{transient sequence} that is executed
speculatively and discarded eventually. %
Since the program must have a secret-independent architectural
observation, there must be no instruction in the main sequence that
depends on the secret. %
However, instructions in the transient sequence (in a Spectre-style
attack these typically belong to a privileged process called by the
attacker) can access the secret but cannot commit. %

In the following, we assume that the transient instructions are
discarded due to a misprediction event. %
For the case that the transient instructions are discarded due to an
exception (Meltdown-style attacks) or for the case that other kinds of
speculations (e.g., load to store dependency prediction) are the cause
of misprediction, the approach is analogous. %
 
For an information leakage to happen, the transient sequence must
affect the behavior of the main sequence, i.e., it must affect the
microarchitectural execution of the program. %
For example, in a \SPECTREXYZ{} attack, the instructions in the
transient sequence affect the timing when instructions commit in the
first block of the main sequence. %
As another example, for the original Spectre attack, the instructions
in the transient sequence create a secret-dependent cache footprint
which induces timing variations in the second block of the main
sequence. %

We observe that a spurious behavior, such as the one described above,
can only lead to a false counterexample to the UPEC proof if it
creates a false interrelation between instructions of the transient
and the main sequence (e.g., a secret value being forwarded from the
transient sequence to the main sequence). %
Due to the UPEC miter structure and the fact that the only difference
between the two instances is the value of the secret, any spurious
behavior within each block is irrelevant for the proof and cannot
produce a false \LALERT{}. %
The reason is that within the main sequence, no execution order can
create a secret-dependent result and within the transient sequence,
none of the results can be committed. %

This means that the bookkeeping mechanisms must be constrained to
ensure the program order and corresponding operand dependencies only
between the three code blocks in Fig.~\ref{fig:transient-inst-seq},
but not necessarily within each block. %
This can be achieved by enforcing that the instructions in the main
sequence do not use the result of transient instructions as an
operand. %
Note that this is a valid constraint because the main sequence block
before the branch never reads an operand from instructions after the
branch instruction, and the main sequence block after the branch never
reads a result from a transient (and as such always discarded)
instruction. %

This observation is key and allows us to approximate microequivalence
effectively by using the branch instruction to split the ROB into the
slots of the main sequence and the slots of the transient sequence. %

Using this partitioning, false operand dependencies between the blocks
can be avoided without formulating complex invariants for source and
destination registers of each in-flight instruction. %
For example, to avoid the false operand dependence for the ADD
instruction of the above example, the proof must be constrained such
that if an ADD instruction from the main sequence (based on its ROB
slot) is being executed by the ALU in both SoC instances, its operands
must be independent of the secret. %
This element of the microequivalence specification is expressed in
Eq.~\ref{eq:me-example} using a pseudo-code notation. %
The approach is
similar for other functional units. %
\begin{align}
\label{eq:me-example}
      (ALU_1.instr\_ID &= ALU_2.instr\_ID) \: \wedge\\
        ALU_1.instr\_ID \: \text{\textit{is assigned} } &  \text{\textit{to main sequence ROB slot}} \nonumber \\ 
\rightarrow 
  (ALU_1.instr\_operands &= ALU_2.instr\_operands) \nonumber
\end{align} 

A more detailed description on how to specify microequivalence for any
processor with a reorder buffer, as well as how to deal with more advanced
features such as nested speculation and load-store dependency speculation, is
available in~\cite{2020-FadihehMueller.etal, web-upec-template}. %

\subsection{UPEC for Processors with Dynamic Register Mapping}

ISAs usually define a set of logical registers (or general-purpose
registers) which are used as source and destination of different
instructions. %
In in-order pipelines, these logical registers are statically mapped
to a set of physical registers. %
They are the architectural (program-visible) registers of the
design. %
However, in order to gain performance improvement by register
renaming, some OOO microarchitectures (e.g.,
BOOM~\cite{2017-CelioChiu.etal}) feature \emph{dynamic register
  mapping}. %
The microarchitecture implements a physical register file, which
usually has more registers than specified in the ISA, as well as a map
table, which holds, at each time point of program execution, the
mapping between logical ISA registers and physical registers. %
This allows for storing both speculative and committed values in the
same register file. %
Such an architecture creates an additional challenge for the UPEC
approach since the proof methodology relies on making the distinction
between \LALERTS{} and \PALERTS{}. %
This means that UPEC must know the architectural state of the system
which relies on the state of the ROB and the map tables. %

Fortunately, there is an efficient solution to this problem. %
In our proof methodology for OOO processors, microequivalence
constrains the search to only transient execution attacks, i.e., UPEC
analyzes windows of the execution where the microarchitectural
observation is independent of the secret. %
This means that the same sequence of values is committed to the
architectural registers in both SoC instances and only the timing (but
not the values) of instruction commit can be secret-dependent. %
This allows us to modify the definition of an \LALERT{} to be used in
Alg.~\ref{alg:upec-base} (lines~7,~19) and Alg.~\ref{alg:upec-step}
(line~6). %
Instead of comparing values of architectural registers, as in
Def.~\ref{def:l-alert}, we check, as given by
Eq.~\ref{eq:lalert-boom-def}, whether or not the instructions in the
two SoC instances reach the head of the ROBs simultaneously. %
\begin{equation}
\label{eq:lalert-boom-def}
   \texttt{\footnotesize soc1.core.rob.head} = \texttt{\footnotesize soc2.core.rob.head} 
\end{equation}

\section{Experiments}
\label{sec:experiments}

The proposed UPEC approach has been evaluated by verifying three
different SoC designs: %
RocketChip (v1.2.0)~\cite{2016-AsanovicAvizienis.etal}, the Berkeley
Out-Of-Order Machine (BOOM v2.0.1)~\cite{2017-CelioChiu.etal} and
Ariane (v4.1.2)~\cite{web-ariane}. %
 All results were
obtained using the commercial property checker OneSpin~360~DV running
on an Intel Core~i7 at 3.4\,GHz, with 32\,GB of RAM. %

\subsection{UPEC for In-order Pipelines}

We evaluated the effectiveness of UPEC for capturing vulnerabilities
in in-order pipelines by targeting different design variants of
RocketChip and the Ariane design. %
The considered RocketChip design variants included the original design
as well as two modified design variants vulnerable to two different
versions of the Orc attack. %
In the first modified variant, we conditionally bypassed one buffer in
the cache-to-core interface pipeline to make the design vulnerable to
the attack described in Sec.~\ref{sec:orc}. %
For the second vulnerable design, RocketChip was modified such that a
cache line refill is not canceled in case of an invalid access. %
While the illegal access itself is not successful but raises an
exception, the cache content is modified and can be analyzed by an
attacker. %

It should be noted that these design modifications are minor (compared
to the overall design size) and they represent a realistic design
optimization. %
The modified designs successfully pass all tests in the suite provided
by the \RISCV{} framework. %

In the following experiments, the secret data is assumed to be in a
protected location, $A$, in the main memory, possibly with a copy in
the data/instruction cache. %
In our RocketChip experiments, memory protection was implemented using
the \textit{Physical Memory Protection (PMP)} scheme of the \RISCV{}
ISA by configuring the memory region holding the location~$A$ of the
secret data as inaccessible in user mode. %
In our Ariane experiment, secret protection is achieved through
virtual address translation, i.e., the memory page holding the
location~$A$ is marked as unreadable and unexecutable by the page
table. %
In our experiments, we assumed correctness of the page table walking
mechanism, i.e., the content of the Translation Lookaside Buffer (TLB)
is always correct w.r.t.\ the page table. %

For RocketChip and Ariane the proof methodology is based on the
properties of Figures~\ref{fig:upec-property-parametric}
and~\ref{fig:upec-inductive-step} and does not employ
\emph{microequivalence}. %
Consequently, these experiments target information leakage through
both functional bugs and transient execution attacks. %
UPEC successfully captured all vulnerabilities to Orc Attacks in the
RocketChip design variants. %
In addition, UPEC found a design bug introducing an ISA incompliance
in the locking mechanism of the \emph{Physical Memory Protection}
(PMP) unit of the original RocketChip design. %
The bug allowed modification of PMP configurations even when the
\textit{lock} bit is set. %
This can compromise security~\cite{2021-MuellerFadiheh.etal} and is
therefore forbidden by the \RISCV{} ISA. %
The vulnerability is removed in the current version of the design. %

UPEC proved the Ariane design to be free of vulnerabilities to
transient execution attacks. %
However, UPEC found an invalid information flow which has relevant
security implications. %
(This was reported to the Ariane development team.) %
The instruction cache, in a certain scenario, allows a user-level
process to refill a cache line with an inaccessible address. %
This may further expand the ability of the attacker to launch
classical side channel attacks, even in cases where there is no page
sharing between attacker and victim. %
We patched the design and re-verified with UPEC. %

In an additional experiment for Ariane we verified the confidentiality
of the page table. %
Protecting physical addresses that are stored in the page table is
important to limit the impact of attack scenarios such as
RowHammer~\cite{2014-KimDaly.etal}. %
We verified the absence of any vulnerability that reveals the physical
address to a user-level process using a modified experimental setup in
which the confidential information of the miter model is an arbitrary
physical page number stored in the TLB. %

\begin{table}
  \centering
  \caption{Computational Effort for Verifying RocketChip and Ariane}
  \label{tab:inorder-experiments}
  \begin{tabular}{l@{}c@{}rrcc}
    \hline
    \hline
    \rule{0pt}{2.2ex}%
    Design variant &
    Proof status &
    CPU &
    Mem. &
           $k$ &
    \#\,P-loc. \\
    
    \hline
    \rule{0pt}{2.2ex}%
    RocketChip modif. 1 &
    \textcolor{red}{detected}  &
    3 min &
    2.7 GB &
             4 &
    \\

    RocketChip modif. 2 &
    \textcolor{red}{detected}  &
    18 min &
    4.6 GB &
             9 &
    \\

    RocketChip original &
    \textcolor{red}{detected}  &
    5 min &
    2.7 GB &
             5  &
    \\

    RocketChip patched &
    \textcolor{darkgreen}{verified}  &
    5 min &
    2.7 GB &
             4 &
                 14 \\

    \hline

    \rule{0pt}{2.2ex}%
    Ariane Original &
    \textcolor{red}{detected}  &
    6 min &
    7.2 GB &
             2 &
    \\

    Ariane Patched &
    \textcolor{darkgreen}{verified}  &
    1 min &
    4 GB &
           2 &
    2 \\

    \hline

    \rule{0pt}{2.2ex}%
    Ariane (page table conf.) &
    \textcolor{darkgreen}{verified}  &
    22 min &
    8.8 GB &
             1 &
    0 \\

    \hline
    \hline
  \end{tabular}
  \vspace{-2ex}
\end{table}

Tab.~\ref{tab:inorder-experiments} shows the proof complexity of the
experiments. %
The designs are verified using the iterative UPEC methodology,
described in Sec.~\ref{sec:upec-iterative}. %
In case of the detected vulnerabilities, the CPU time, memory
consumption and property time window ($k$) of the proof that returned
the corresponding \LALERT{} is reported. %
All vulnerabilities were found within the \emph{UPEC\_Induction\_Base}
algorithm, before proceeding to \emph{UPEC\_Induction\_Step}. %
In case of the designs verified to be secure, the reported CPU time,
memory consumption and property time window ($k$) corresponds to the
most complex proof in terms of run time within the
\emph{UPEC\_Induction\_Base} and \emph{UPEC\_Induction\_Step}
algorithms. %
It should be noted that in all these experiments, the most complex
proof was always within the last iteration of \emph{UPEC\_Induction\_Base}. %
  For the secure designs, the last column in
  Tab.~\ref{tab:inorder-experiments} shows the number of \PLOCS{} that
  have to be proven secure in \emph{UPEC\_Induction\_Step}. %

In these experiments, it took approximately 10 person-days of manual
work to verify each design, which was predominantly spent in analyzing
the provided counterexamples and debugging the design. %
Considering the complexity of the designs, the incurred manual effort
is small compared to the efforts required for the functional
verification of these processors. %

\begin{table}
  \centering
  \caption{Proof Complexity in different settings}
  \label{tab:complexity-comparison}
  \begin{tabular}{lrr}
    \hline
    \hline
    \rule{0pt}{2.2ex}%
     &
    CPU &
    Mem. \\
    \hline
    \rule{0pt}{2.2ex}%
    32-bit datapath &
    5 min &
    2.7 GB \\

    64-bit datapath &
    16 min &
    6 GB \\

    \hline

    \rule{0pt}{2.2ex}%
    UPEC w. cone-of-influence reduction &
    5 min &
    2.7 GB  \\

    UPEC w/o. cone-of-influence reduction &
    61 min &
    6 GB \\

    \hline
    \hline
  \end{tabular}
  \vspace{-2ex}
\end{table}

Tab.~\ref{tab:complexity-comparison} evaluates the effect of different
factors on the proof complexity, based on the UPEC experiment on
RocketChip. %
Both 32-bit and~64-bit versions of the design are verified and the
highest proof runtime is reported, which shows the robustness of our
approach w.r.t.\ datapath complexity. %
The table also shows the significant improvements that can be achieved
by simplifying the %
  \PROVE{} part of the property %
  based on cone-of-influence
reduction (cf.~Sec~\ref{sec:cone-of-influence-reduction}). %

\subsection{UPEC for OOO Pipelines}

The BOOM processor is of particular interest in our experiments since
it is known to be vulnerable to Spectre-style attacks while deemed
secure with respect to Meltdown. %
It is a full-grown SoC with an OOO-core which features a branch
prediction unit with support for nested branches, virtual memory
translation with a TLB, %
a non-blocking data cache with miss status handling registers (MSHR),
a page table walker, a physical register file with dynamic register
mapping and other features typically employed with OOO-cores. %
  BOOM's performance is comparable to ARM cores between Cortex A9 and
  A15, depending on its configuration. %
The verified Boom design (single core and peripherals) consists of
more than 650\,k state bits. %
The only difference with the previous experiments is that the UPEC
proofs were constrained using the concept of
\emph{microequivalence}~(Sec.~\ref{sec:upec-for-ooo}). %

As a demonstration that we can focus separately on different classes
of attacks, in our experiments, we decomposed the proofs into checks
for Meltdown versus Spectre, i.e., as defined in
Sec.~\ref{sec:upec-theoretical}, we assume that the instruction
accessing the secret either has the proper privilege to do so (Spectre
class) or not (Meltdown class). %
Furthermore, the secret is assumed to reside in the main memory, with
the possibility of a copy in the data cache. %

Tab.~\ref{tab:results} has three sections, describing our experiments
with the checks for the Spectre and Meltdown classes, as well as for
page table confidentiality. %
Similar to Tab.~\ref{tab:inorder-experiments}, it shows the
computational effort for the most complex proofs within
\emph{UPEC\_Induction\_Base} and \emph{UPEC\_Induction\_Step} %
for our patch-and-verify flow, as described below. %
It should be noted that the vulnerabilities were detected by
\LALERT{}s in \emph{UPEC\_Induction\_Base}. %
  For the secure designs, the last column in the table shows the
  number of \PLOCS{} that had to be proven secure in
  \emph{UPEC\_Induction\_Step}. %

Tab.~\ref{tab:manual-work} shows the amount of manual work required to
develop and prove the UPEC property for BOOM. %

\begin{table}
  \centering
  \caption{Computational Effort for Verifying BOOM}
  \label{tab:results}
  \begin{tabular}{l@{}clcccc}
    \hline
    \hline
    \rule{0pt}{2.2ex}%
    Check &
    Patch &
    Proof result &
    CPU  &
    Mem.  &
    $k$ &
    \# P-
    \\
     & no. & & (min) & (GB) & & loc. \\
     
    \hline
    \rule{0pt}{2.2ex}%
    Spectre &
    0 &
    \textcolor{red}{\ATTACK{} v1} &
        2  &
      9  &
    5  & \\

    & 1 & 
    \textcolor{red}{\ATTACK{} v2} &
    16  &
    14  &
    7 & \\

    & 2 & 
    \textcolor{red}{Spectre (cache)} &
    14  &
    14  &
    7 & \\

    & 3 & 
    \textcolor{darkgreen}{secure} &
    5  &
    6  &
    2 & 0\\

    \hline

    \rule{0pt}{2.2ex}%
    Meltdown & 0 &
    \textcolor{darkgreen}{secure} &
    5  &
    6  &
    2 & 3\\

    \hline

    \rule{0pt}{2.2ex}%
    Page table conf. &
    0 & 
    \textcolor{darkgreen}{secure} &
    5  &
    5.9  &
    1 & 3\\
    \hline
    \hline
  \end{tabular}
  \vspace{-2ex}
\end{table}

For the class of Spectre attacks, our approach generated
counterexamples to the UPEC property in terms of \LALERTS{}
(cf.~Sec.~\ref{sec:upec-bounded-model}) demonstrating that the
considered BOOM design is vulnerable to \SPECTREXYZ{}, a so far
unknown variant of Spectre, described in Sec.~\ref{sec:spectre-stc}. %
We employed an iterative design procedure where we iteratively patch
the vulnerability detected in a selected UPEC counterexample, and then
re-verify the design using UPEC. %
After fixing the \SPECTREXYZ{} vulnerability through a minor fix in
the first iteration, UPEC identified (by \LALERT{}) a second version
of the \SPECTREXYZ{} vulnerability. %
This version is similar to the first one %
except that port contention now happens on the TLB rather than on the
write port to the register file. %
The patch-and-verify flow can be repeated until all the
vulnerabilities have been removed. %

In the third iteration of our flow, we picked a UPEC counterexample
pointing to the original Spectre attack, which uses the cache as side
channel~\cite{2018-KocherGenkin.etal}. %
We applied a simple and rather conservative patch for this
vulnerability. %
Our fix for Spectre prevents speculative load instructions to execute
before all preceding branch instructions are resolved. %
The patched design coming out of %
this third iteration (``secure design variant'' in
Tab.~\ref{tab:results}) was then %
formally verified to be secure with respect to transient execution
attacks. %
Although the fix incurs performance penalties, it still allows for
speculative execution of the majority of instructions. %
 
In a separate experiment, we also examined the original design for the
class of Meltdown attacks, as defined in
Sec.~\ref{sec:upec-theoretical}. %
The computational effort for this experiment, which proves security
 of the original BOOM~v2.0.1 w.r.t.\ Meltdown,
is also listed in Tab.~\ref{tab:results}. %

To the best of our knowledge, presently no other method is capable of solving the above RTL verification tasks. %
As a result of our experiments, the patched BOOM design is the first
formally proven RTL model of a secure speculative execution
processor. %

\begin{table}
  \centering
  \caption{Required Manual Effort to Develop and Prove UPEC Property}
  \label{tab:manual-work}
  \begin{tabular}{ll}
    \hline
    \hline
    \rule{0pt}{2.7ex}Task&
    Manual effort \\
    \hline
    \rule{0pt}{2.7ex}Microequivalence &
    4 person-weeks \\
    Invariants &
    4 person-days \\
    Property and miter &
    2 person-hours \\
    \hline
    \hline
  \end{tabular}
  \vspace{-3ex}
\end{table}

\section{Conclusion}
\label{sec:conclusion}

The paper has shown that transient execution side channels can occur
also in simple processors. %
They do not only result from certain architectural features, such as
out-of-order execution and speculative execution, but can also be inserted by low-level RTL design modifications. %
Such vulnerabilities may be introduced inadvertently by seemingly
innocuous design decisions and, even worse, they may also be
deliberately created by a malicious SW/HW provider to create an
invisible ``backdoor''. %
Such a new kind of Trojan can be extremely hard to find, since it does
not conform to the conventional Trojan models (trigger and payload),
and, more importantly, it neither corrupts the functionality of the
design nor does it add any redundant logic. %
With the increasing role of shared HW and SW infrastructures involving
components from numerous providers as well as open-source domains,
such a risk must be given appropriate attention. %

We have demonstrated that UPEC is capable of detecting TESs in an
exhaustive, scalable and highly automated way. %
This is a significant improvement over the state of the art,
especially since also vulnerabilities can be detected which are based
on so far unknown exfiltration channels. %

The required manual work in UPEC is small compared to the total SoC
design and verification effort at the RTL. %
The bulk of this effort is spent on analyzing the provided
counterexamples and security violations. %

The proposed approach, due to its exhaustiveness, has promise for an
efficient solution regarding legacy HW. %
By analyzing the RTL design of legacy systems the UPEC user can
collect all possible attack scenarios feasible in the design. %
The collected information can be passed on to the SW domain to enforce
SW fixes only for the necessary cases and relevant gadgets. %
Future work will explore how UPEC can support methodologies employing
contracts between HW and SW~\cite{2020-GuarnieriKoepf.etal}
such that a sound compositionality can be achieved between measures at
the SW level and the RTL. %

\ifCLASSOPTIONcaptionsoff
  \newpage
\fi

\bibliographystyle{IEEEtran}
\bibliography{refs3}

\vspace{-8ex}
\begin{IEEEbiography}[{\includegraphics[width=1in,height=1.25in,clip,keepaspectratio]{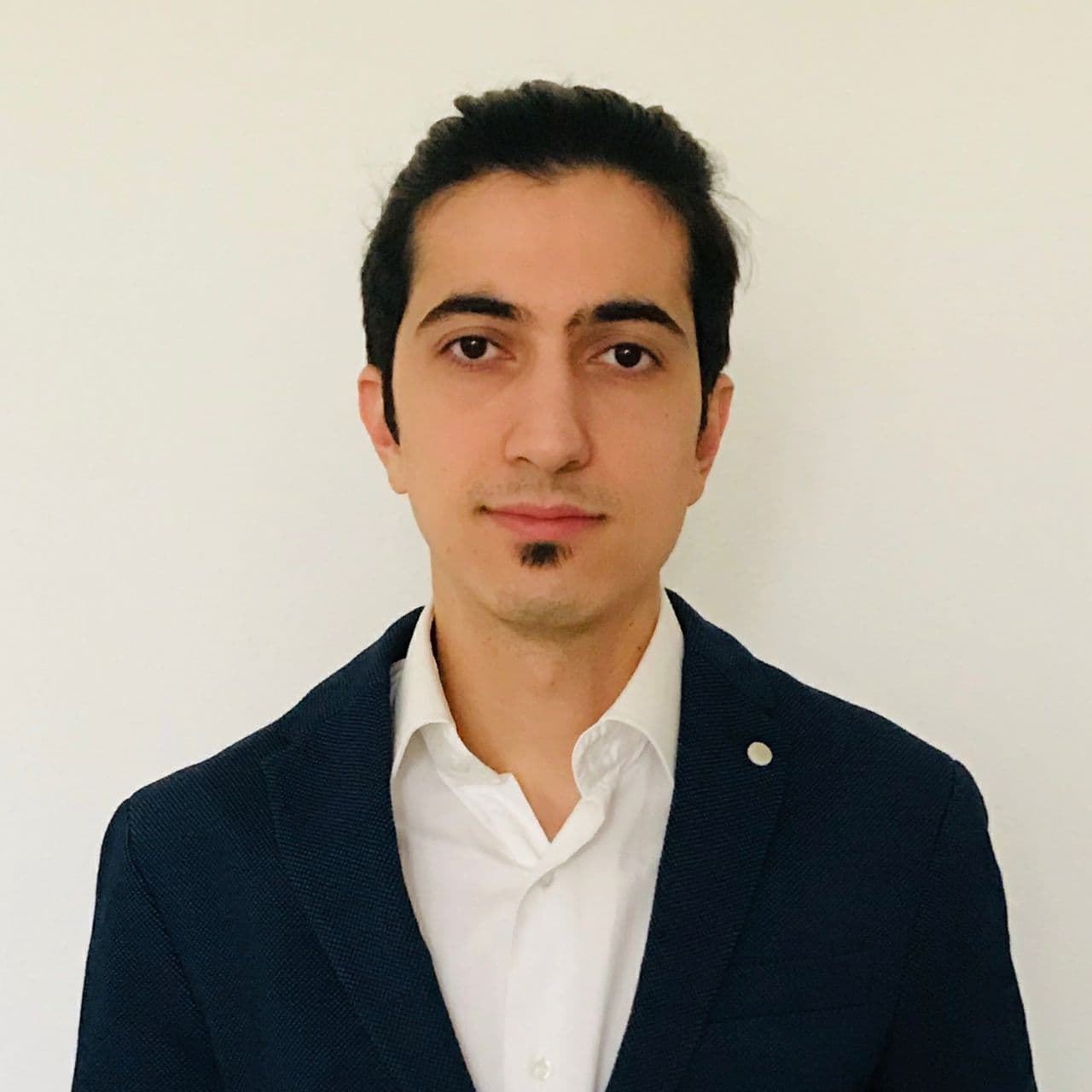}}]{Mohammad Rahmani Fadiheh}
  received the M.Sc.\ %
  degree in Electrical and Computer Engineering from Technische
  Universit\"at Kaiserslautern, Kaiserslautern, Germany, in 2017. %
  He is currently pursuing a doctoral degree with Electronic Design
  Automation group at the same institution. %
  His current research interests include formal verification, hardware
  security, side channel attacks and secure hardware design. %
  \vspace{-10ex}
\end{IEEEbiography}

\begin{IEEEbiography}[{\includegraphics[width=1in,height=1.25in,clip,keepaspectratio]{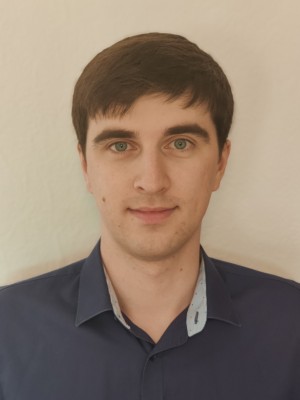}}]{Alex Wezel}
  received the M.Sc.\ %
  degree in Electrical and Computer Engineering from Technische
  Universit\"at Kaiserslautern, Kaiserslautern, Germany, in 2020. %
  He is currently pursuing a doctoral degree with Electronic Design
  Automation group at the same institution. %
  His current research interests include formal verification, hardware
  security and methodologies for fixing vulnerable hardware designs. %
  \vspace{-10ex}
\end{IEEEbiography}

\begin{IEEEbiography}[{\includegraphics[width=1in,height=1.25in,clip,keepaspectratio]{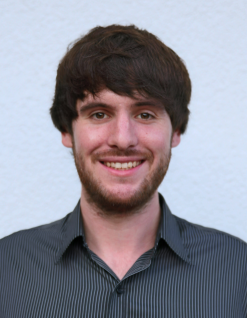}}]{Johannes
    M\"uller}
  received the Dipl.-Ing.\ %
  degree in Electrical and Computer Engineering from Technische
  Universit\"at Kaiserslautern, Kaiserslautern, Germany, in 2018. %
  He is currently pursuing a doctoral degree with Electronic Design
  Automation group at the same institution. %
  His current research interests include formal verification, access
  control in SoCs and microarchitectural timing side channels. %
  \vspace{-10ex}
\end{IEEEbiography}

\begin{IEEEbiography}[{\includegraphics[width=1in,height=1.25in,clip,keepaspectratio]{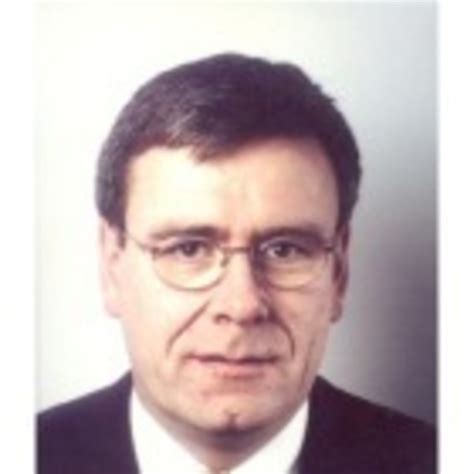}}]{J\"org Bormann}
  received a Master's degree in Mathematics from Technische
  Universit\"at Karlsruhe, Germany, in 1990, and the Dr.-Ing. %
  degree from Technische Universit\"at Kaiserslautern, Germany, in
  2009. %
  He is with Siemens EDA, where he works on advanced formal
  verification topics. %
  His research interests encompass easily usable correctness proofs like
  GapFree Verification, formal security verification and integration
  of formal HW verification and system development. %
  \vspace{-10ex}
\end{IEEEbiography}

\begin{IEEEbiography}[{\includegraphics[width=1in,height=1.25in,clip,keepaspectratio]{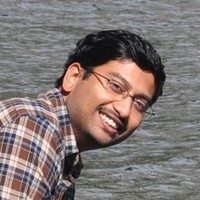}}]{Sayak Ray}
  received the B.Sc. %
  degree in Computer Science and Engineering from Jadavpur University,
  Jadavpur, India, in 2004, and the M.Sc.\ %
  degree in Computer Science and Engineering from Indian Institute of
  Technology, Kharagpur, India, in 2007, and the Ph.D.\ %
  degree in Electrical Engineering and Computer Science from
  University of California Berkeley, Berkeley, USA, in 2013. %
  He is currently with Intel Product Assurance and Security, Intel
  Corporation, USA. %
  \vspace{-10ex}
\end{IEEEbiography}

\begin{IEEEbiography}[{\includegraphics[width=1in,height=1.25in,clip,keepaspectratio]{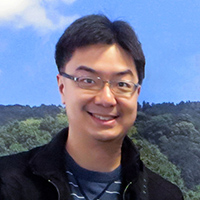}}]{Jason M. %
Fung}
received the first B.S. %
degree in Computer Science and Mathematics and the second B.S. %
degree in Electrical and Computer Engineering from Carnegie Mellon
University, Pittsburgh, PA, USA, in 1997, and the M.S.degree in ECE
from Carnegie Mellon University in 1998. %
He is the Director of Offensive Security Research and Academic
Research Engagement at Intel Product Assurance and Security. %
\vspace{-10ex}
\end{IEEEbiography}

\begin{IEEEbiography}[{\includegraphics[width=1in,height=1.25in,clip,keepaspectratio]{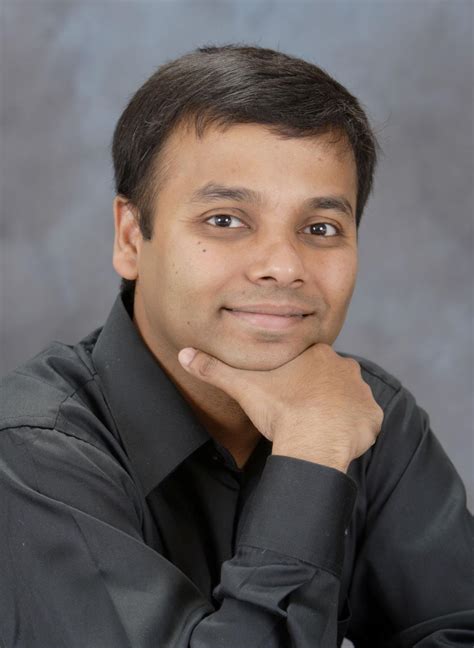}}]{Subhasish Mitra}
  is a Professor of Electrical Engineering and of Computer Science at
  Stanford University, Stanford, CA, USA, where he directs the
  Stanford Robust Systems Group, co-leads the computation focus area
  of the Stanford SystemX Alliance, and is also a Faculty Member of
  the Stanford Neurosciences Institute. %
  He holds the Carnot Chair of Excellence in Nanosystems with
  CEA-LETI, Grenoble, France. %
  His current research interests range broadly across robust
  computing, nanosystems, very large-scale integration (VLSI) design,
  validation, test and electronic design automation, and
  neurosciences. %
  \vspace{-10ex}

\end{IEEEbiography}

\begin{IEEEbiography}[{\includegraphics[width=1in,height=1.25in,clip,keepaspectratio]{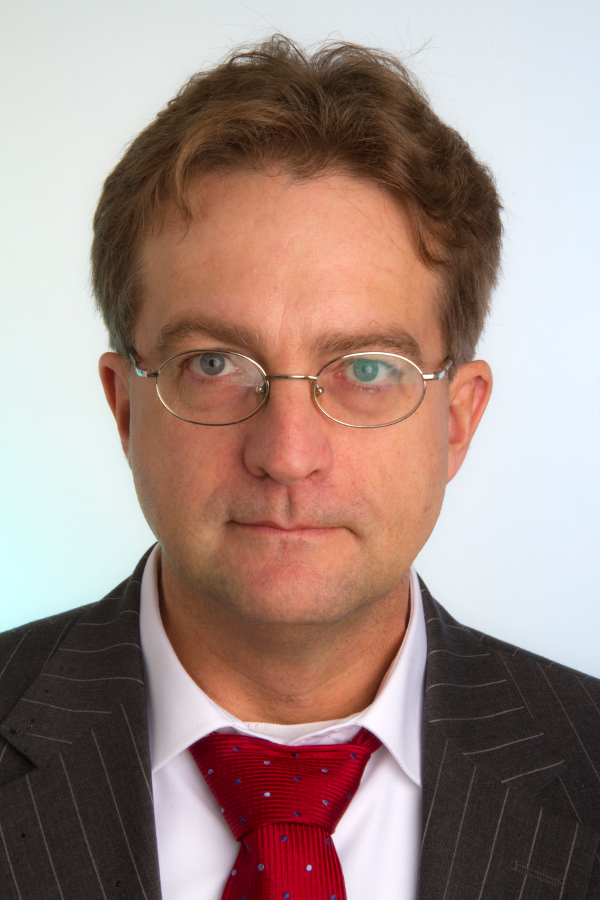}}]{Dominik Stoffel}
  received the Dipl.-Ing.\ %
  degree from the University of Karlsruhe, Karlsruhe, Germany, in
  1992, and the Dr.phil.nat degree from Johann-Wolfgang Goethe
  Universit\"at, Frankfurt, Germany, in 1999. %
  He has held positions with Mercedes-Benz, Stuttgart, Germany, and
  the Max-Planck Fault-Tolerant Computing Group, Potsdam, Germany. %
  Since 2001, he has been a Research Scientist and a Professor with
  the Electronic Design Automation group, Technische Universit\"at
  Kaiserslautern, Kaiserslautern, Germany. %
  His current research interests include design and verification
  methodologies for Systems-on-Chip. %
  \vspace{-10ex}
\end{IEEEbiography}

\begin{IEEEbiography}[{\includegraphics[width=1in,height=1.25in,clip,keepaspectratio]{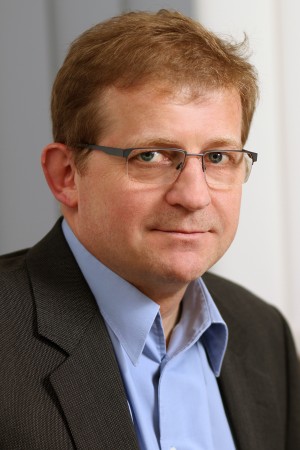}}]{Wolfgang Kunz}
  received the Dipl.-Ing.\ %
  degree in Electrical Engineering from University of Karlsruhe in
  1989, the Dr.-Ing. %
  degree in Electrical Engineering from University of Hannover in 1992
  and the Habilitation degree from University of Potsdam (Max Planck
  Society, Fault-Tolerant Computing Group) in 1996. %
  Since 2001 he is a professor at the Department of Electrical \&
  Computer Engineering at Technische Universit\"at Kaiserslautern,
  Germany. %

  Wolfgang Kunz conducts research in the area of System-on-Chip design
  and verification. %
  His current research interests include verification-driven design
  methodologies for hardware and firmware, safety analysis and
  hardware security. %

\end{IEEEbiography}

\end{document}